\documentclass{aa}
\usepackage[varg]{txfonts}
\usepackage{natbib}
\bibpunct{(}{)}{;}{a}{}{,}
\begin{document}

\title{Probing X-ray burst -- accretion disk interaction in low mass X-ray binaries through kilohertz quasiperiodic oscillations} 
\author{P. Peille
  \and J-F. Olive 
     \and D. Barret} 
\institute{Universit\'e de Toulouse, UPS-OMP, IRAP, Toulouse, France 
\\ and CNRS Institut de Recherche en Astrophysique et Plan\'etologie, 9 Av. colonel Roche, BP 44346, F-31028 Toulouse cedex 4, France\\
\email{ppeille@irap.omp.eu}}
\date{Received 8 March 2014 / Accepted 24 May 2014}
\abstract{The intense radiation flux of Type I X-ray bursts is expected to interact with the accretion flow around neutron stars. High frequency quasiperiodic oscillations (kHz QPOs), observed at frequencies matching orbital frequencies at tens of gravitational radii, offer a unique probe of the innermost disk regions. In this paper, we follow the lower kHz QPOs, in response to Type I X-ray bursts, in two prototypical QPO sources, namely 4U 1636-536 and 4U 1608-522, as observed by the Proportional Counter Array of the \emph{Rossi} X-ray Timing Explorer. We have selected a sample of 15 bursts for which the kHz QPO frequency can be tracked on timescales commensurable with the burst durations (tens of seconds). We find evidence that the QPOs are affected for over $\sim$200 s during one exceptionally long burst and $\sim$100 s during two others (although at a less significant level), while the burst emission has already decayed to a level that would enable the pre-burst QPO to be detected. On the other hand, for most of our burst-kHz QPO sample, we show that the QPO is detected as soon as the statistics allow and in the best cases, we are able to set an upper limit of $\sim$20 s on the recovery time of the QPO. This diversity of behavior cannot be related to differences in burst peak luminosity.
We discuss these results in the framework of recent findings that accretion onto the neutron star may be enhanced during Type I X-ray bursts. The subsequent disk depletion could explain the disappearance of the QPO for $\sim$100 s, as possibly observed in two events. However, alternative scenarios would have to be invoked for explaining the short recovery timescales inferred from most bursts. Heating of the innermost disk regions would be a possibility, although we cannot exclude that the burst does not affect the QPO emission at all. Clearly the combination of fast timing and spectral information of Type I X-ray bursts holds great potential in the study of the dynamics of the inner accretion flow around neutron stars. However, as we show, breakthrough observations will require a timing instrument providing at least ten times the effective area of the RXTE/PCA.}

\keywords{accretion, accretion disks -- X-rays: bursts -- stars: individual: 4U 1636-536 -- stars: individual: 4U 1608-522 --\\X-rays: binaries}

\titlerunning{Probing X-ray burst -- accretion disk interaction through kHz QPOs}

\maketitle

\section{Introduction}
Illumination of accreting disks during Type I X-ray bursts gives us the opportunity to study the innermost regions of the accretion flow around neutron stars \citep{Galloway:2008, Strohmayer:2003aa,Cumming:2004aa}. Evidence of an interaction between the burst emission and the inner disk has been reported in a few individual bursts \citep{Yu:1999,Kuulkers:2003,Chen:2011,int-Zand:2011,Serino:2012,Degenaar:2013}. Disk depletion through radiation drag, heating of the inner disk, and even radiatively or thermally powered winds were discussed by \citet{Ballantyne:2004} and \citet{Ballantyne:2005}, who attempted to explain the time evolution of the properties of the \emph{superburst} of \object{4U\,1820-303}. More recently, \citet{Worpel:2013aa}, fitting the spectra of all photospheric radius expansion (PRE) bursts observed with the \emph{Rossi} X-ray Timing Explorer (RXTE) as the sum of a blackbody and a scalable continuum having the shape of the pre-burst persistent emission, reported a systematic increase of the persistent emission during the burst. They interpreted this result as evidence of an accretion rate enhancement due to a rapid increase of the radiation torque on a thin accretion disk, as formalized early on by \citet{Walker:1992}. A similar finding was reported by \citet{int-Zand:2013aa} for a burst observed simultaneously by \emph{Chandra} and RXTE, although disk reprocessing of the burst emission was preferred as an alternative explanation. 
\begin{table*}
	\caption{Principal characteristics of the selected X-ray bursts.}
	\label{tab:liste_bursts}
	\centering
	\begin{tabular}{c c c c c c c c c}
	\hline\hline
	Source & Burst ID & Obs ID & Start time & $L_{\text{peak}}$\tablefootmark{a}  & $E_{\text{tot}}$\tablefootmark{a}  & $\tau$\tablefootmark{a}  & $C_{\text{pers}}$ & $C_{\text{peak}}$\\
	& & & (RXTE time) & (10$^{38}$ ergs/s) & (10$^{39}$ ergs) & (s) & (cts/s) & (cts/s)\\
	\hline
	4U 1636-536 & 4 & 10088-01-08-030 & 94671416 & 2.68 &  5.54 & 20.7 & 788 & 9931\\
	 & 6 & 30053-02-02-02 & 146144682 & 2.89 & 1.78 & 6.2 & 679 & 9740\\
	 & 9 & 40028-01-02-00 & 162722852 & 2.94 & 1.84 & 6.3 & 569 & 9927\\
	 & 21 & 40028-01-18-000 & 208401523 & 2.79 & 2.35 & 7.0 & 737 & 9916\\
	 & 22 & 40028-01-18-00 & 208429016 & 2.76 & 1.85 & 6.7 & 574 & 10000\\
	 & 23 & 40028-01-19-00 & 208740744 & 2.79 & 2.13 & 7.6 & 541 & 10116\\
	 & 39 & 60032-01-09-00 & 242286131 & 1.46 & 1.33 & 9.1 & 417 & 7089\\
	 & 40 & 60032-01-09-01 & 242295310 & 1.06 & 1.14 & 10.9 & 425 & 5355\\
	 & 41 & 60032-01-10-00 & 242708641 & 2.00 & 1.50 & 7.5 & 428 & 9611\\
	 & 168 & 91024-01-30-10 & 374626248 & 2.98 & 2.06 & 6.9 & 473 & 9604\\
	\hline
	4U 1608-522 & 4 & 30062-02-01-000 & 133389809 & 1.43 & 1.35 & 9.1 & 643 & 9004\\
	& 5 & 30062-01-01-00 & 133625123 & 2.34 & 2.32 & 9.9 & 543 & 10339\\
	& 21 & 70059-01-20-00 & 273983179 & 2.33 & 1.93 & 8.3 & 617 & 10562\\
	& 23 & 70059-03-01-000 & 274421898 & 2.39 & 3.16 & 13.2 & 517 & 10549\\
	& 24 & 70059-03-01-000 & 274434678 & 1.08 & 1.1 & 10.2 & 510 & 9528\\
	\hline
	\end{tabular}
	\tablefoot{Burst ID corresponds to the burst number in the \citet{Galloway:2008} burst catalog, Obs ID to the identification number of the RXTE observation in which the burst can be found, Start time to the RXTE time of the burst peak used to define the date $t = 0$ in Sect.~\ref{data_analysis}, $L_{\text{peak}}$ to the peak luminosity, $E_{\text{tot}}$ to the total energy, $\tau$ to the decay time constant, $C_{\text{pers}}$ to the raw persistent count rate before the burst, and $C_{\text{peak}}$ to its value at burst peak. \tablefoottext{a}{Values taken from \citet{Galloway:2008}. Distances of 6 and 3.6~kpc were taken to compute the peak luminosity $L_{\text{peak}}$ and total energy} $E_{\text{tot}}$ for 4U 1636-536 and 4U 1608-522, respectively \citep{Pandel:2008,Nakamura:1989}. }
\end{table*}

In this paper, we follow a different path and look at kilohertz quasiperiodic oscillations (kHz QPOs) in response to Type I X-ray bursts (see \citealt{van-der-Klis:2006} for a review of kHz QPOs). Although there is not yet a consensus on the origin of kHz QPOs, it is generally agreed that they arise from the vicinity of the neutron star \citep[e.g.,][]{Barret:2013}, most likely in a region to be exposed to the burst emission. It is therefore worth investigating how the QPO properties react to X-ray bursts. For this purpose,  we use the RXTE archival data of  \object{4U 1636-536} and \object{4U 1608-522}, known as frequent bursters \citep{Galloway:2008} and whose lower kHz QPOs can be detected and followed on the burst duration timescales of tens of seconds \citep{Mendez:1999,Barret:2005,Barret:2006}. 
\section{Data analysis}
\label{data_analysis}

We retrieved the archived data of the Proportional Counter Array (PCA) onboard RXTE for all the observations containing an X-ray burst for both 4U 1636-536 and 4U 1608-522, using the catalog provided by \citet{Galloway:2008}. In total, we obtained 172 bursts for 4U 1636-536 and 31 for 4U 1608-522, respectively. Using the Science Events of the PCA, we computed 1/2$^{12}$~s resolution light curves from $-$200 to +400 s with respect to the burst peak (whenever possible, i.e., when no observation gap restricted this time interval). The date $t = 0$ was fixed for each burst at the peak count rate using 1 s resolution light curves. In order to limit the influence of the nonmodulated burst photons in our analysis, while keeping enough statistics for the QPO detection, we chose to restrict our study to the 6--50 keV band. The burst emission dominates in the soft X-ray band while the QPO modulated photons have a harder spectrum \citep{Berger:1996}.

	\subsection{Data reduction}
	\label{data_reduction}

Dynamical Fourier power density spectra (PDS) containing the time series of Leahy normalized PDS \citep{Leahy:1983} were computed. These spectra were obtained with an integration time of 1 s and a Nyquist frequency of $f_{Ny}$ = 2048 Hz. Segments of data containing an X-ray burst and a QPO  detectable on short timescales ($\sim$20--30~s) were identified. A list of 10 and 5 bursts in 4U 1636-536 and 4U 1608-522, respectively, was so obtained (see Table \ref{tab:liste_bursts} for a summary of the main properties of these bursts).

	\subsection{Initial results}

\begin{figure}[!t]
	\resizebox{\hsize}{!}{\includegraphics{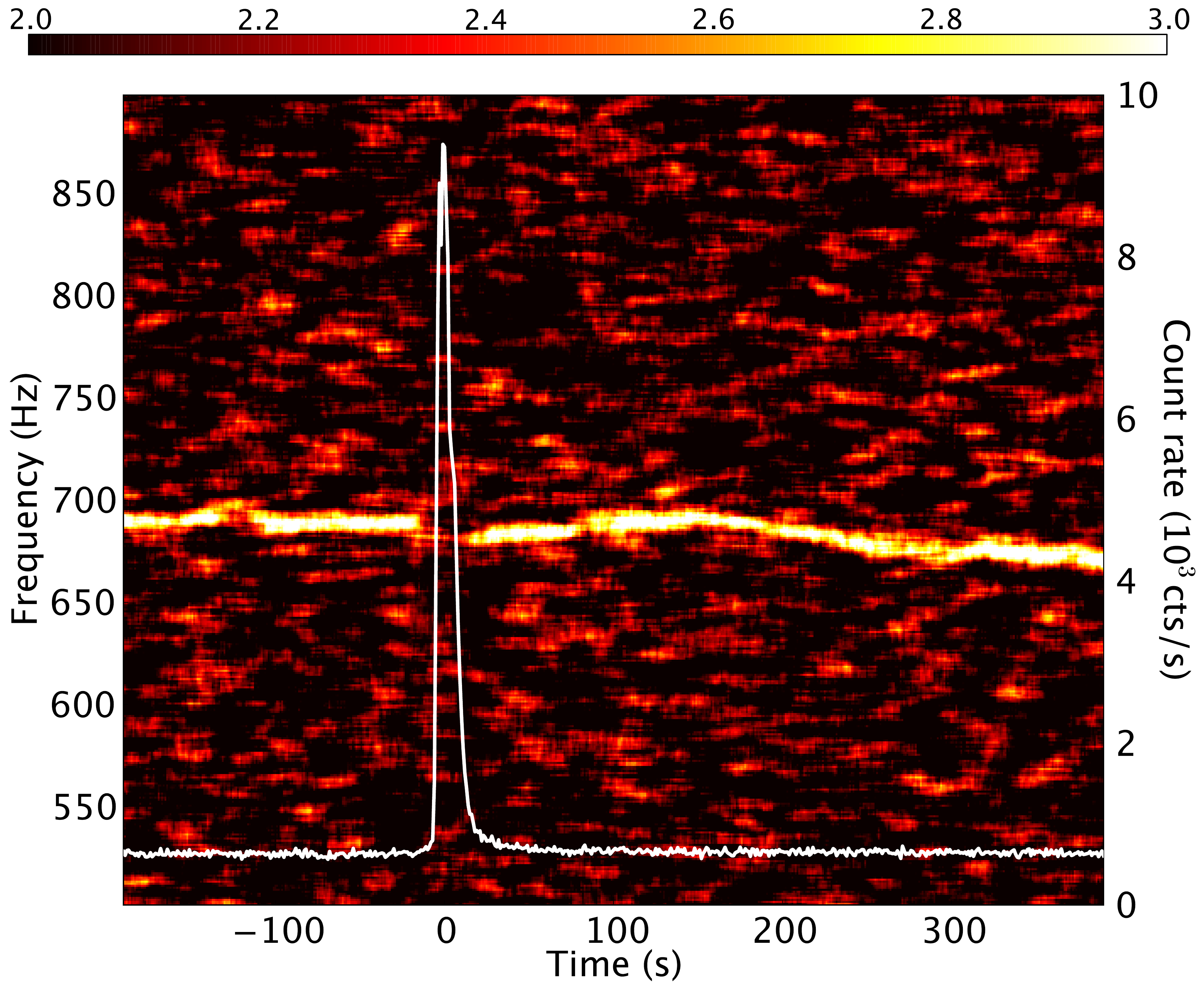}}
	\centering
	\caption{Dynamical PDS and light curve (white line) around burst 4 of 4U 1608-52. The image corresponds to a series of 1~s PDS plotted as a function of time, convolved with a 6~Hz and 20~s averaging kernel for better visibility. Power is color coded with a linear scale between 2 (black) and 3 (white).}
	\label{fig:disp}
\end{figure}

\begin{figure*}
	\centering
	\begin{tabular}{ccc}
		\includegraphics[width = 0.315\linewidth]{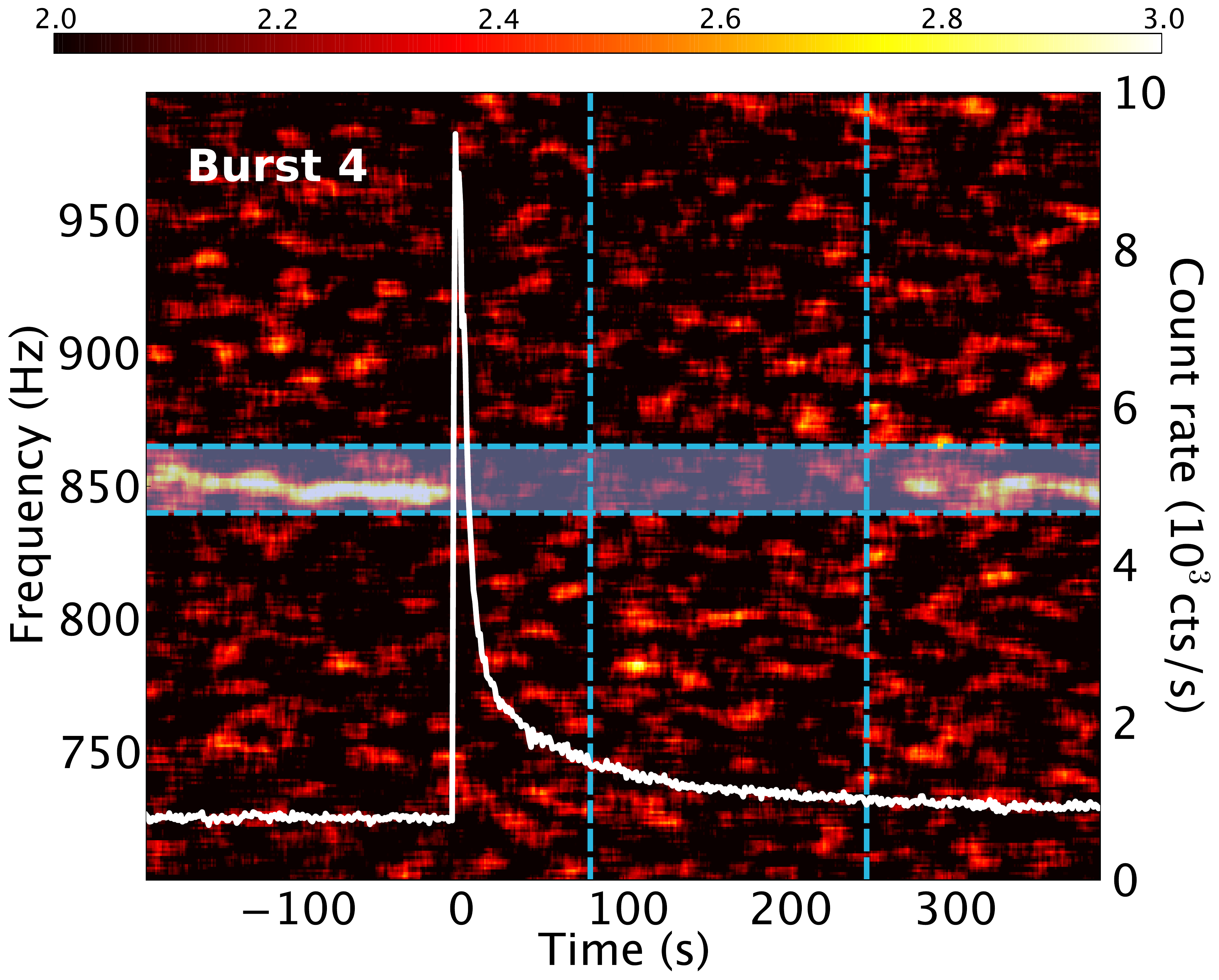} &
		\includegraphics[width = 0.315\linewidth]{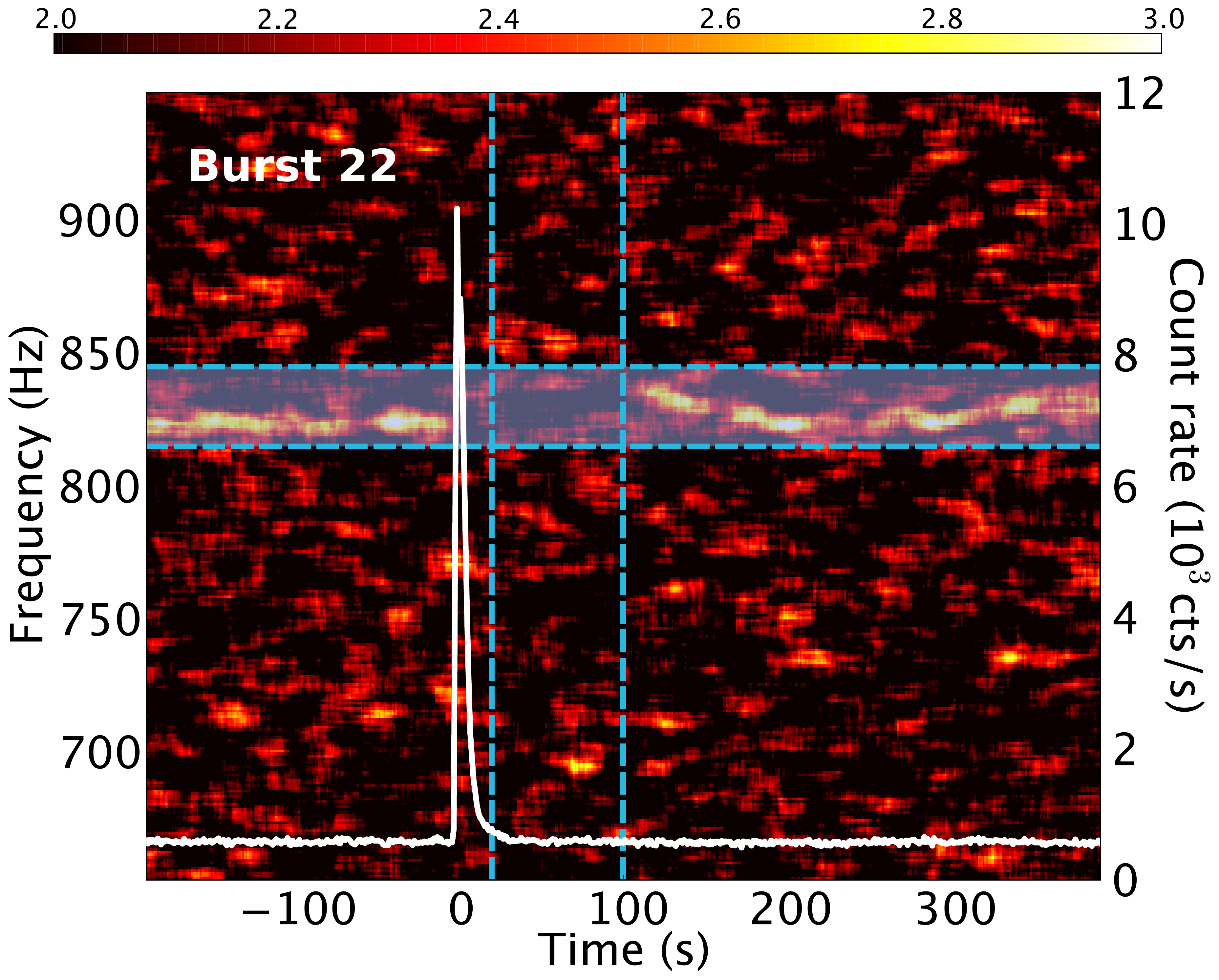} &
		\includegraphics[width = 0.315\linewidth]{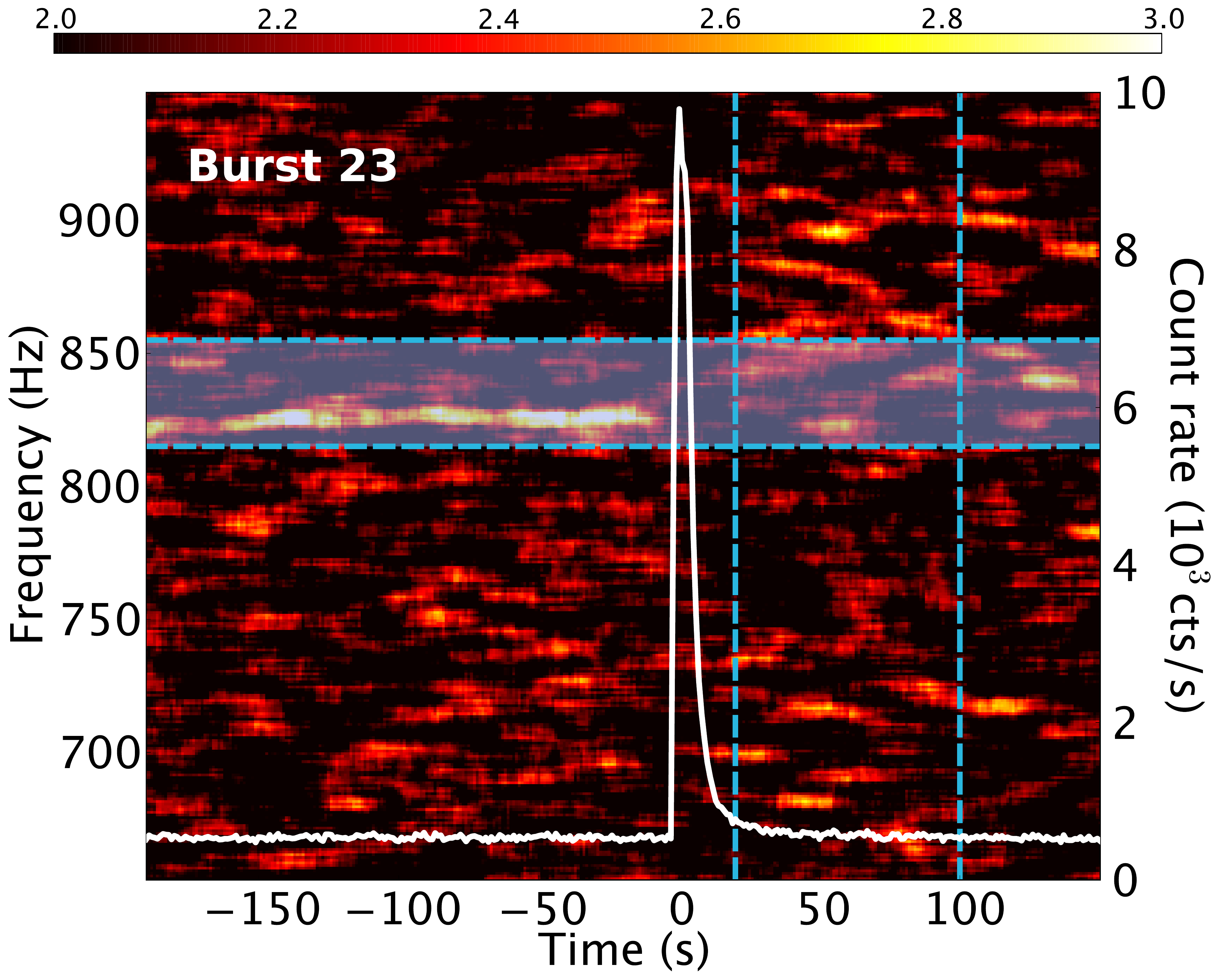}		
	\end{tabular}
	\caption{Dynamical PDS and light curves (white lines) around bursts 4 (left), 22 (middle), and 23 (right) of 4U 1636-536. The image corresponds to a series of 1~s PDS plotted as a function of time, convolved with a 6~Hz and 20~s averaging kernel for better visibility. Power is color coded with a linear scale between 2 (black) and 3 (white). The blue zones and vertical dashed lines represent the time and frequency intervals used to probe the significance of the QPO nondetection in the simulations, namely 80$-$248~s and 840$-$865~Hz for burst 4, 20$-$100~s and 815$-$845~Hz for burst 22, and 20$-$100~s and 815$-$855~Hz for burst 23. We note that for burst 23 an observation interruption limits the analysis to $\sim$150~s after the burst.}
	\label{fig:4_22_23}
\end{figure*}

The most common pattern observed in the dynamical PDS is illustrated in Fig. \ref{fig:disp}: the QPO is clearly detected both before and after the X-ray burst at about the same frequency (i.e., at a frequency consistent with the usual drift of the QPO frequency), but in an interval of $\sim$20--30~s after the onset, the QPO is not detected. This nondetection can be explained by the addition of the nonmodulated burst photons\footnote{The significance of excess power is proportional to $\frac{S^2}{S+B}\text{rms}^2$. $S$, $B$, and rms correspond to the signal count rate, background count rate, and to the fractional rms amplitude of the QPO emission \citep{van-der-Klis:1989}. A burst corresponds to a large increase of $B$, causing the significance of the signal to drop abruptly.}. This pattern is, however, not observed in three bursts of 4U 1636-536, for which the QPO remains undetected for up to several hundreds of seconds, while the overall source emission has returned to the level it had before the burst (see Fig. \ref{fig:4_22_23}).

\begin{table*}
	\caption{kHz QPO shapes before and after the X-ray bursts.}
	\label{tab:para_fit}
	\centering
\begin{tabular}{c c c c c c c c c c}
    \hline\hline
    Source & Burst ID & Freq$_{\text{pers}}$ & $R_{\text{pers}}$ & rms$_\text{pers}$ (\%) & $w_{\text{pers}}$ (Hz) & Freq$_{\text{burst}}$ & $R_{\text{burst}}$  & rms$_{\text{burst}}$ (\%) & $w_{\text{burst}}$ (Hz)\\
    \hline
    4U 1636-536 & 4 & 848 & 8.5$\pm$0.4 & 10.4$\pm$0.2 & 3.5$\pm$0.3 & (u.l.) & 1.2 (u.l.) & 4.9 (u.l.) & (u.l.)\\
    & 6 & 874 & 6.7$\pm$0.4 & 10.0$\pm$0.3 & 4.3$\pm$0.4 & 876$\pm$2.5 & 7.9$\pm$2.8 & 10.9$\pm$1.9 & 13.8$\pm$6.3\\
    & 9 & 862 & 8.2$\pm$0.4 & 11.9$\pm$0.3 & 5.5$\pm$0.4 & 855$\pm$0.6 & 5.3$\pm$1.9 & 9.8$\pm$1.8 & 4.8$\pm$3.0\\
    & 21 & 859 & 9.5$\pm$0.5 & 11.4$\pm$0.3 & 5.6$\pm$0.5 & 854$\pm$0.4 & 7.2$\pm$1.3 & 10.0$\pm$0.9 & 3.5$\pm$0.9\\
    & 22 & 823 & 5.9$\pm$0.4 & 10.3$\pm$0.3 & 3.9$\pm$0.3 & (u.l.) & 2.6 (u.l.) & 6.9 (u.l.) & (u.l.)\\
    & 23 & 825 & 5.4$\pm$0.3 & 10.0$\pm$0.3 & 2.2$\pm$0.2 & (u.l.) & 2.6 (u.l.) & 7.1 (u.l.) & (u.l.)\\
    & 39 & 802 & 3.2$\pm$0.3 & 8.7$\pm$0.4 & 2.0$\pm$0.4 & 796$\pm$2.1 & 5.1$\pm$1.8 & 11.3$\pm$2.0 & 8.3$\pm$3.8\\
    & 40 & 843 & 5.2$\pm$0.4 & 11.0$\pm$0.4 & 4.2$\pm$0.5 & 839$\pm$0.5 & 5.7$\pm$1.3 & 11.9$\pm$1.4 & 3.7$\pm$1.3\\
    & 41 & 820 & 3.9$\pm$0.3 & 9.5$\pm$0.4 & 2.7$\pm$0.3 & 826$\pm$0.7 & 5.5$\pm$1.4 & 12.0$\pm$1.5 & 5.0$\pm$1.7\\
    & 168 & 849 & 4.9$\pm$0.3 & 10.0$\pm$0.3 & 2.7$\pm$0.3 & 849$\pm$0.7 & 6.7$\pm$1.5 & 12.2$\pm$1.4 & 5.3$\pm$1.6\\
    \hline
    4U 1608-522 & 4 & 689 & 10.9$\pm$0.5 & 12.9$\pm$0.3 & 5.6$\pm$0.3 & 684$\pm$0.5 & 10.9$\pm$1.8 & 13.6$\pm$1.1 & 5.5$\pm$1.3\\
    & 5 & 740 & 8.8$\pm$0.4 & 12.8$\pm$0.3 & 4.7$\pm$0.3 & 722$\pm$1.2 & 11.6$\pm$3.0 & 15.7$\pm$2.0 & 12.1$\pm$4.4\\
    & 21\tablefootmark{a} & 703 & 10.8$\pm$0.4 & 13.2$\pm$0.3 & 4.7$\pm$0.3 & 704$\pm$0.8 & 10.1$\pm$2.0 & 13.4$\pm$1.3 & 7.9$\pm$2.1\\
    & 23 & 813 & 5.8$\pm$0.4 & 10.4$\pm$0.3 & 4.1$\pm$0.3 & 801$\pm$2.8 & 6.0$\pm$1.8 & 12.8$\pm$1.9 & 6.7$\pm$2.8\\
    & 24 & 805 & 8.5$\pm$0.5 & 12.8$\pm$0.4 & 5.7$\pm$0.5 & 790$\pm$0.5 & 9.3$\pm$1.8 & 13.6$\pm$1.3 & 5.3$\pm$1.5\\
    \hline
\end{tabular}
	\tablefoot{Burst ID corresponds to the burst number in the \citet{Galloway:2008} burst catalog, Freq to the centroid frequency of the QPO, $R$ to the Lorentzian normalization factor, rms to the rms amplitude of the QPO, and $w$ to its width. The QPO parameters are given both before (pers index) and after the burst (burst index). The errors here are $1\sigma$ errors. For bursts 4, 22, and 23 of 4U 1636-536, the QPO rms amplitude 90\% confidence upper limits were computed during the nondetection gap, namely 80--248~s for the first nondetection and 20--100~s for the other two using the pre-burst width as well as a frequency fixed to the position of the most significant excess (denoted u.l. in the table). For all other bursts, the QPO parameters after the burst were computed in the same time interval, 20--100~s, after the burst peak.\tablefoottext{a}{For this burst, not enough data were recorded before the burst and the QPO parameters have been estimated away from the burst (200 s after), and taken as representative of the QPO parameters before the burst.}}
\end{table*}

\begin{figure*}[!t]
	\centering
	\begin{tabular}{ccc}
		\includegraphics[width = 0.315\linewidth]{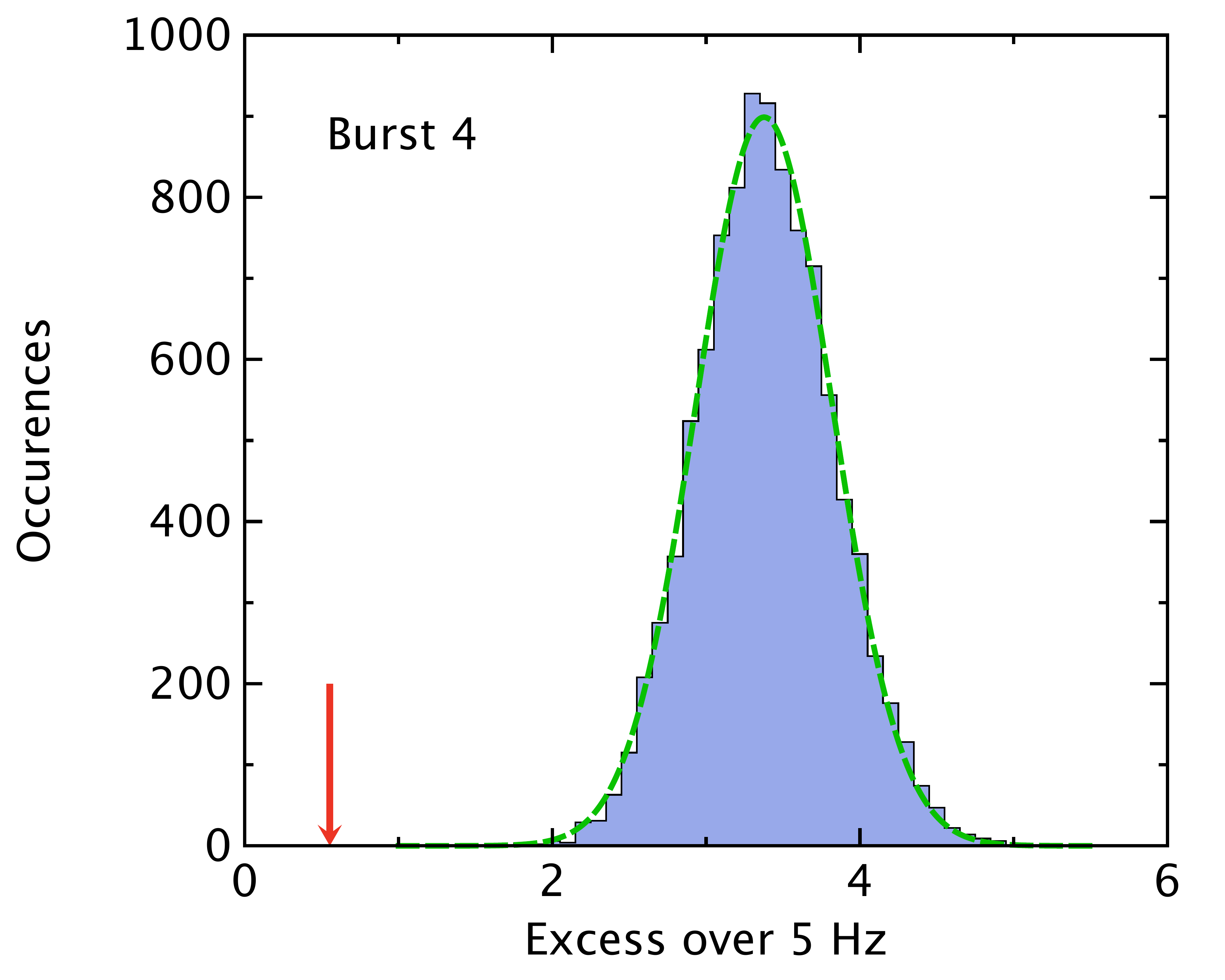} &
		\includegraphics[width = 0.315\linewidth]{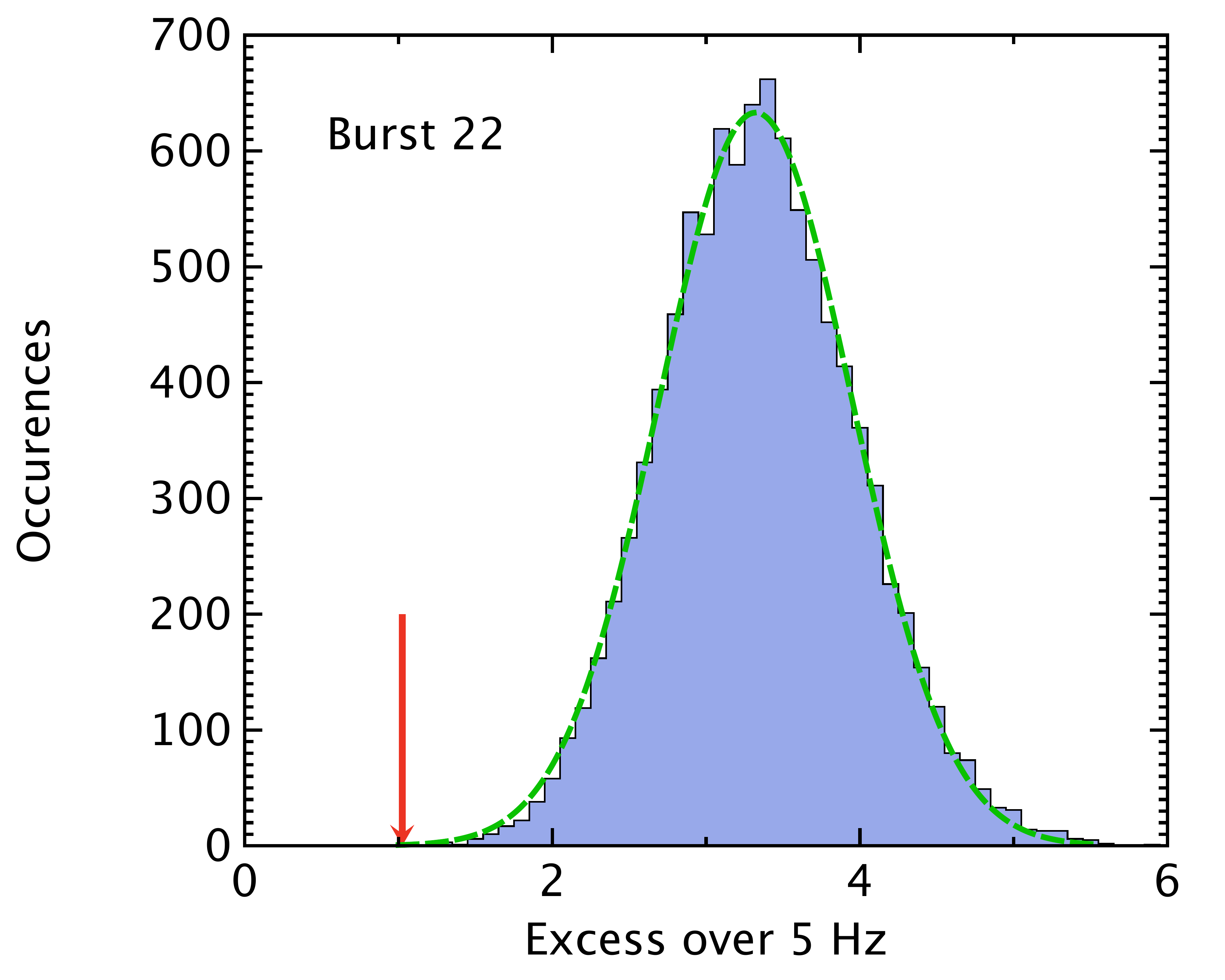} &
		\includegraphics[width = 0.315\linewidth]{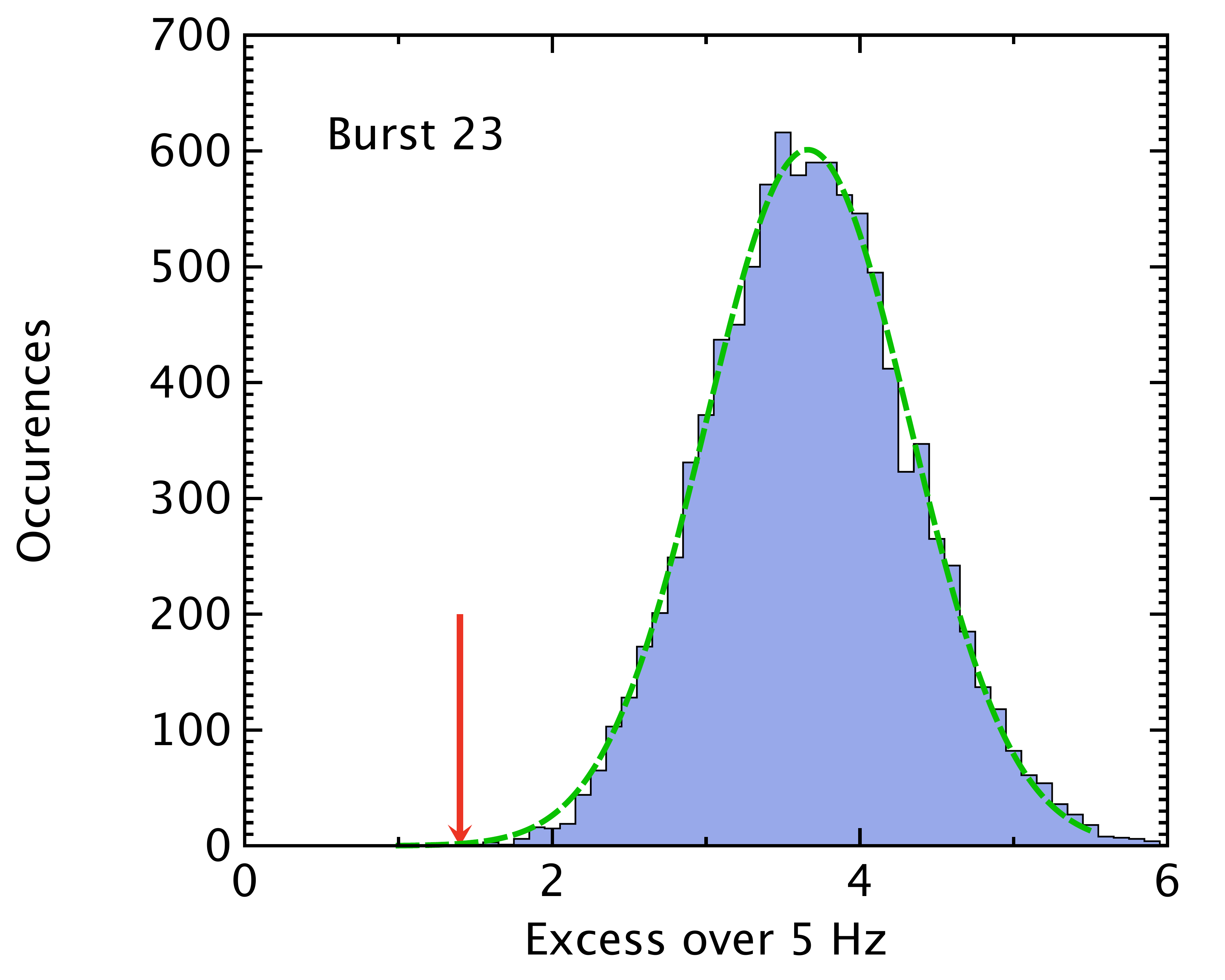}
	\end{tabular}
	\caption{Results of the PDS simulations for bursts 4 (left), 22 (middle), and 23 (right) of 4U 1636-536. The blue histograms correspond to the distribution of the excesses found in the simulations (see text), the red arrows to the values found in the data, and the green dashed lines to Gaussian fits of the distributions.}
	\label{fig:distrib_simu}
\end{figure*}

We then simulated the effect of the burst photons on the QPO detectability. For this purpose, we first estimated the QPO parameters (amplitude, frequency, width) using segments of 100~s duration prior to the bursts between $t=-200~$s and $t=-10~$s, stepped by 10~s (sliding window). The QPO was fitted by a Lorentzian with the maximum likelihood method described in \citet{Barret:2012aa} and was found to be highly significant in each segment. The final parameters were thus obtained by computing the weighted average of the best fit parameters found in each segment (see Table \ref{tab:para_fit}). A persistent emission light curve with the same time resolution as the data (1/2$^{12}$~s) containing such a QPO was then randomly simulated following \citet{Timmer:1995}. The burst light curve was itself computed on the same time resolution, starting from the interpolation of the one recorded with one-second time bins. Poisson noise fluctuations were added to the high-time-resolution burst light curve, which was then added to the light curve of the persistent emission containing the QPO. The summed light curve was then processed the same way as the real data to produce one-second PDS. This procedure was repeated 10 000 times. The PDS were averaged within the nondetection gap for both the data and the simulations, and excesses of power were searched over the same frequency interval around the QPO frequency in a 5~Hz window (see Fig. \ref{fig:4_22_23}). In conditions representative of the three bursts, the simulations systematically reproduced an excess of power larger than the one found in the data, strongly suggesting that the QPO was affected by the burst (see Fig. \ref{fig:distrib_simu}). A more quantitative estimate of the significance of these nondetections can be obtained by comparing the excess of power found in the data with the distributions produced by the simulations. For that particular PDS integration time (1 s), Gaussian fits yield probabilities of observing the gap if the QPO parameters were not affected by the burst of $9.2\times 10^{-11}$ for burst 4, $1.3\times 10^{-4}$ for burst 22, and $3.3\times 10^{-4}$ for burst 23 (see Fig. \ref{fig:distrib_simu}). The nondetection gaps during bursts 22 and 23 can therefore only be considered as marginally significant.

\begin{figure}[!t]
	\resizebox{\hsize}{!}{\includegraphics{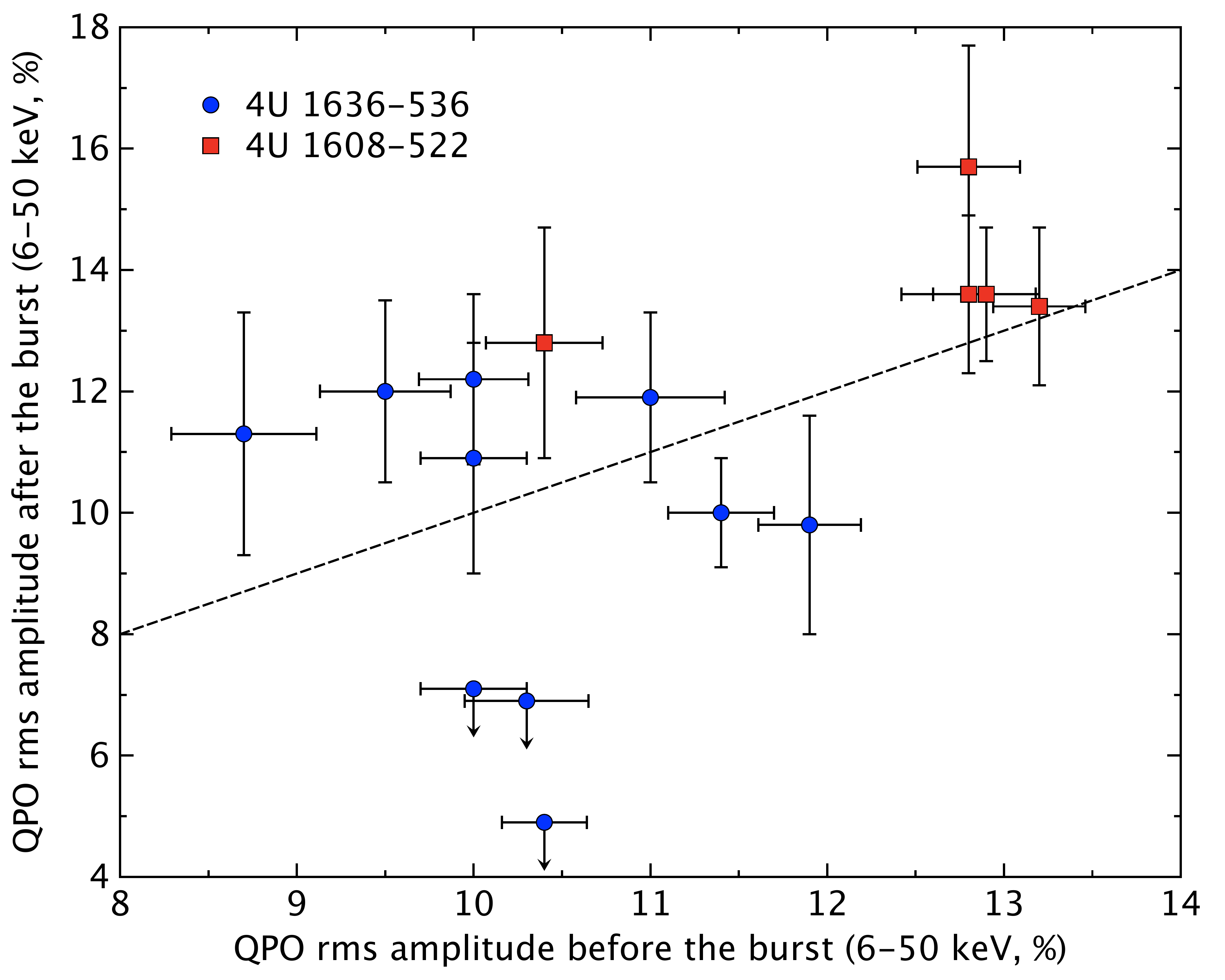}}
	\centering
	\caption{Distribution of the QPO rms amplitude measured after the burst against the one measured before the burst with 1$\sigma$ error bars (see Table~\ref{tab:para_fit}). The upper limits are 90\% level. The blue circles correspond to bursts of 4U 1636-536, the red squares to bursts of 4U 1608-522. The dashed line corresponds to equal amplitudes before and after the burst.}
	\label{fig:rms_bursts}
\end{figure}

Another way to illustrate this is by comparing the rms amplitude of the QPO prior to the burst and the rms upper limit on the QPO during the nondetection gap, considering the QPO width unchanged and the burst photons contributing to the noise. As listed in Table \ref{tab:para_fit}, 90\% confidence level upper limits are significantly below the QPO rms measured prior to the burst. On the other hand, for all the other bursts of 4U 1636-536 (and 4U 1608-522), the rms amplitude of the QPO measured before and after the burst are consistent with one another, as shown in Fig. \ref{fig:rms_bursts} (see Table \ref{tab:para_fit} for more detail). We note that in order to take into account the changing count rate during the burst decay and the fact that burst photons are not modulated, the rms amplitude of the QPO was computed from the Lorentzian normalization factor $R$ \citep{Barret:2012aa} using the relation
\begin{equation}
\text{rms} = \sqrt{\dfrac{T}{C_{\text{pers}}^2\sum_t1/C(t)}R},
\end{equation}
where $C(t)$ is the one-second resolution count rate in the time interval of duration $T$ and $C_{\text{pers}}$ the count rate in the persistent emission.

\begin{figure}[!t]
	\resizebox{\hsize}{!}{\includegraphics{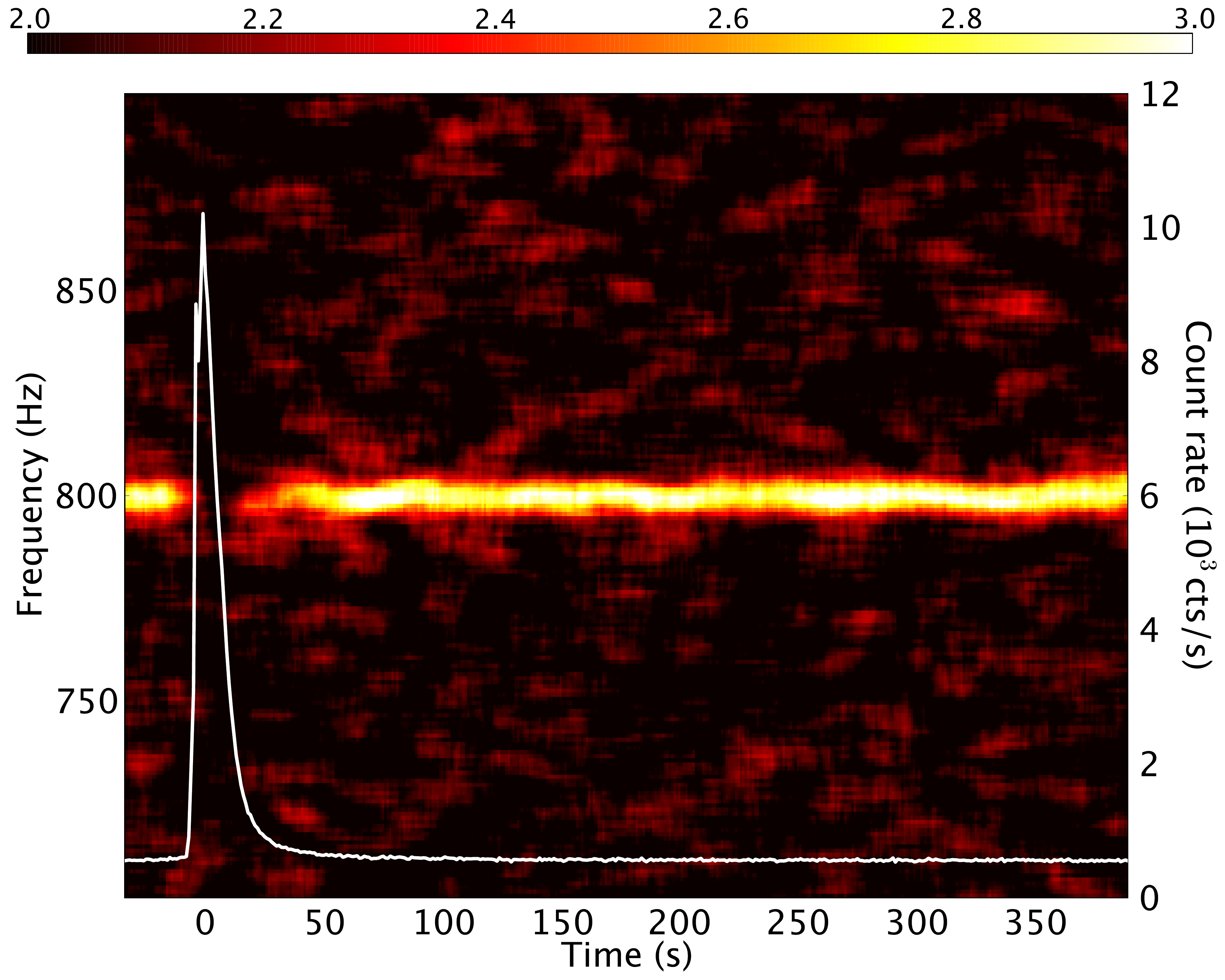}}
	\caption{Dynamical PDS and mean light curve (white line) resulting from the stacking of all the 4U 1608-52 observations. The image corresponds to the succession of 1~s PDS plotted as a function of time, convolved with a 6 Hz and 20 s averaging kernel for better visibility. Power is color-coded with a linear scale between 2\,(black) and 3\,(white). We note that the plot starts at $-$43~s because of an observation gap for burst 21.}
	\label{fig:stack1608}
\end{figure}

\subsection{Stacking of the bursts}

As shown in Fig. \ref{fig:rms_bursts}, all QPOs from 4U 1608-522 show a stable rms amplitude. In addition, for 4U 1608-522, the QPOs are easier to track on short timescales than for 4U 1636-536. As a way to improve our QPO detection sensitivity around the burst, one can combine the PDS from different bursts, after aligning them to a reference frequency. The QPO frequency was tracked before and after the burst using an algorithm similar to the one presented in \citet[][]{Barret:2006}, which uses the centroid of the excess power as a proxy of the QPO frequency (when the significance of the QPO is not sufficient to obtain a stable fit). We note that because no signal can be detected during the bursts, the frequency is obtained by interpolating between the values before and after the bursts. We have checked through extensive Monte Carlo PDS simulations of QPOs of varying frequency, in the presence of a typical X-ray burst of 4U 1608-522, that such an algorithm did not introduce any systematic biases in the recovered QPO parameters, and is therefore suitable for combining PDS of different bursts. The dynamical PDS produced for 4U~1608-522 after all the PDS were aligned to an 800 Hz reference frequency is shown in Fig.~\ref{fig:stack1608}. Clearly the nondetection gap is now restricted to an area very close to the burst peak. For example, a 6.6$\sigma$ excess of power is detected in a 15-second interval, 15~s after the burst peak. We have thus set a conservative upper limit of $\sim$20~s on the QPO reappearance or on its recovery time.

\begin{figure}[!t]
	\centering
	\resizebox{\hsize}{!}{\includegraphics{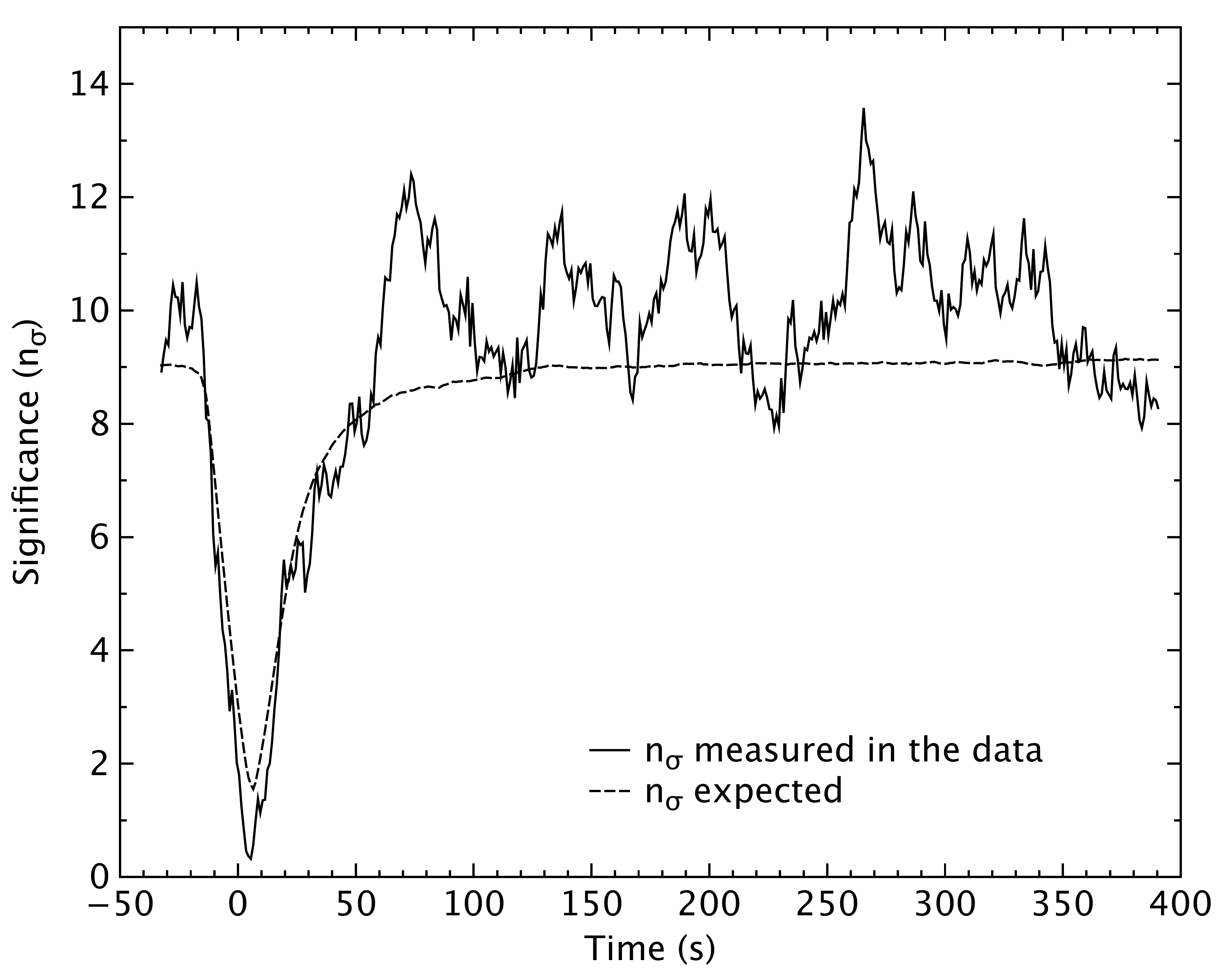}}
	\caption{Significance of the maximum excess over 4~Hz in the 790$-$810~Hz band as a function of time in the dynamical PDS resulting from the stack of the 4U 1608-52 observations (see text). The solid line corresponds to significance values measured in the data while the dashed line is the level expected from the burst light curves and QPO profiles measured before the burst. We note that the plot starts at $-$43~s because of an observation gap for burst 21.}
	\label{fig:excess1608}
\end{figure}

Sliding a 20~s window over the burst (with one-second steps, i.e., steps of one second each), one can track the evolution of the significance of the QPO detection as a function of time, and compare it to the one for which the QPO parameters are unchanged during the bursts and only the burst photons add to the noise to lower the detection significance. This is shown in Fig. \ref{fig:excess1608}; the solid curve shows the evolution of the significance of the maximum power excess over 4~Hz (total power exceeding the Poisson level of 2) measured as a function of time in the stacked PDS between 790~Hz and 810~Hz. The dashed line corresponds to the level expected assuming that the QPO parameters measured before the burst do not change during the bursts. It is remarkable how close the observed and predicted profiles are\footnote{The small difference between the two curves in the persistent emission after the burst is due to the measurement of the power excess used here being sensible to positive noise fluctuations around the QPO profile.}, suggesting that for our sample of 4U 1608-522 bursts, the QPOs were unaffected by the presence of the bursts.  

\section{Comparison with burst parameters}

\begin{figure*}
	\centering
	\begin{tabular}{cc}
	\includegraphics[width = 0.4\linewidth]{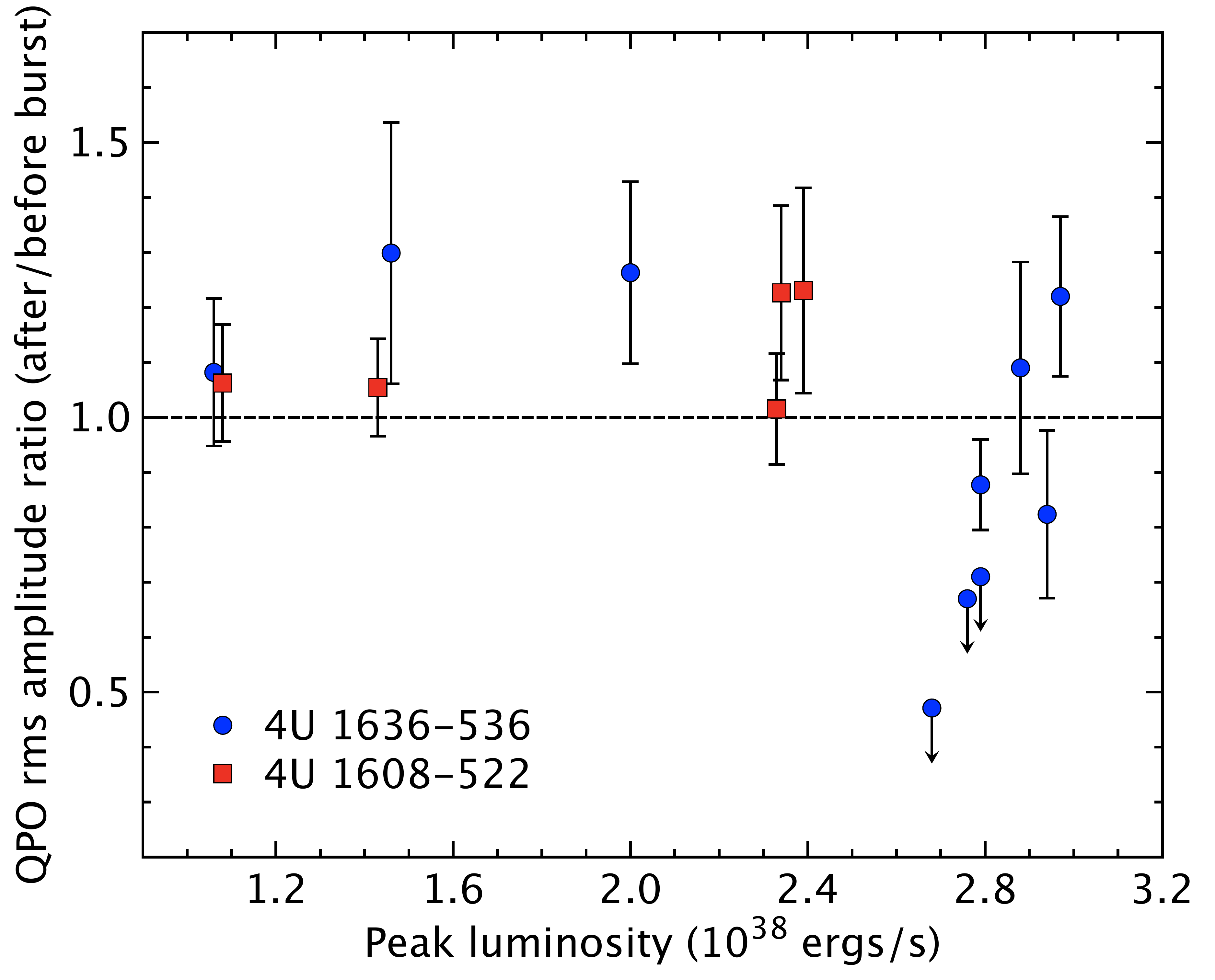} &
	\includegraphics[width = 0.4\linewidth]{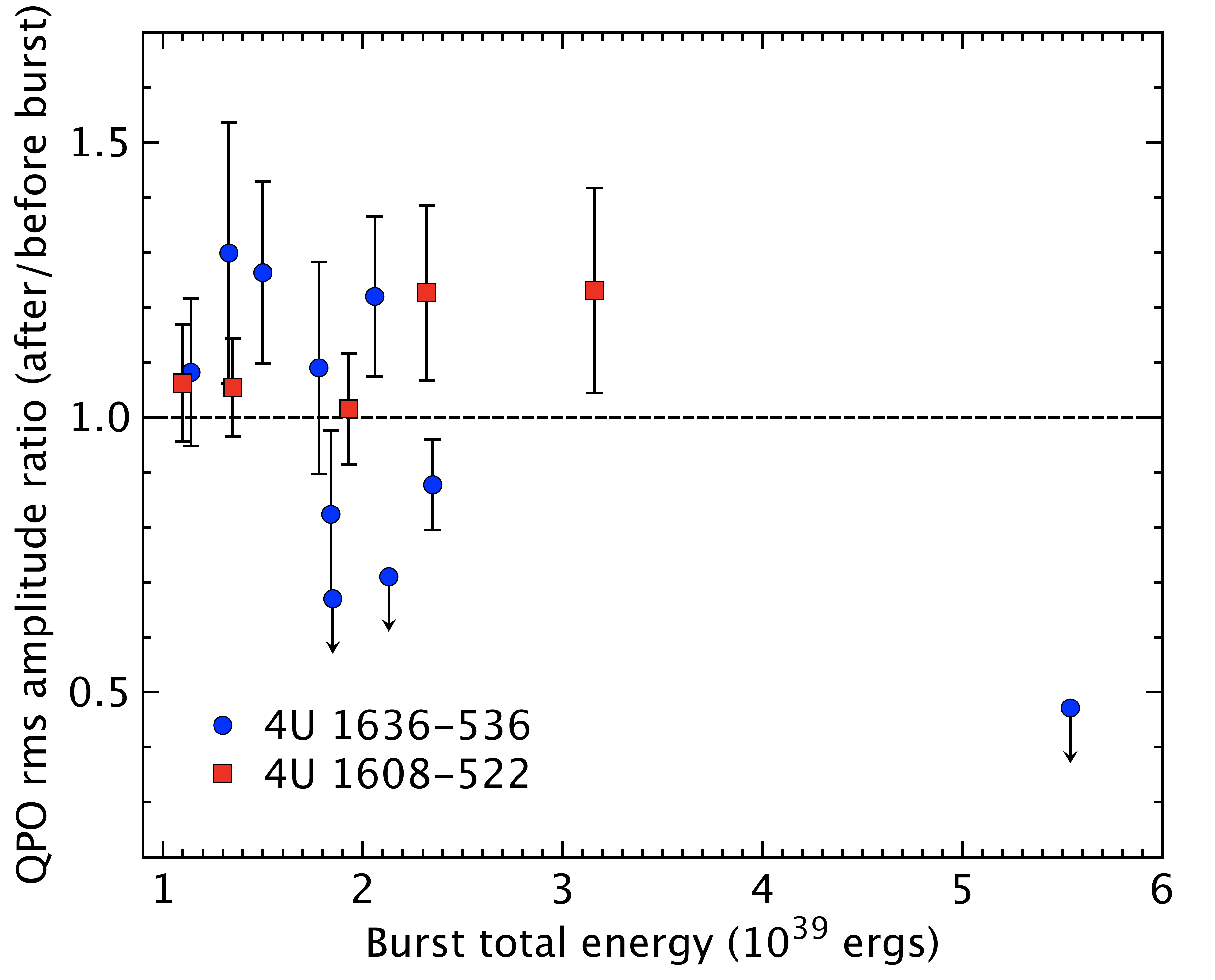}\\
	\includegraphics[width = 0.4\linewidth]{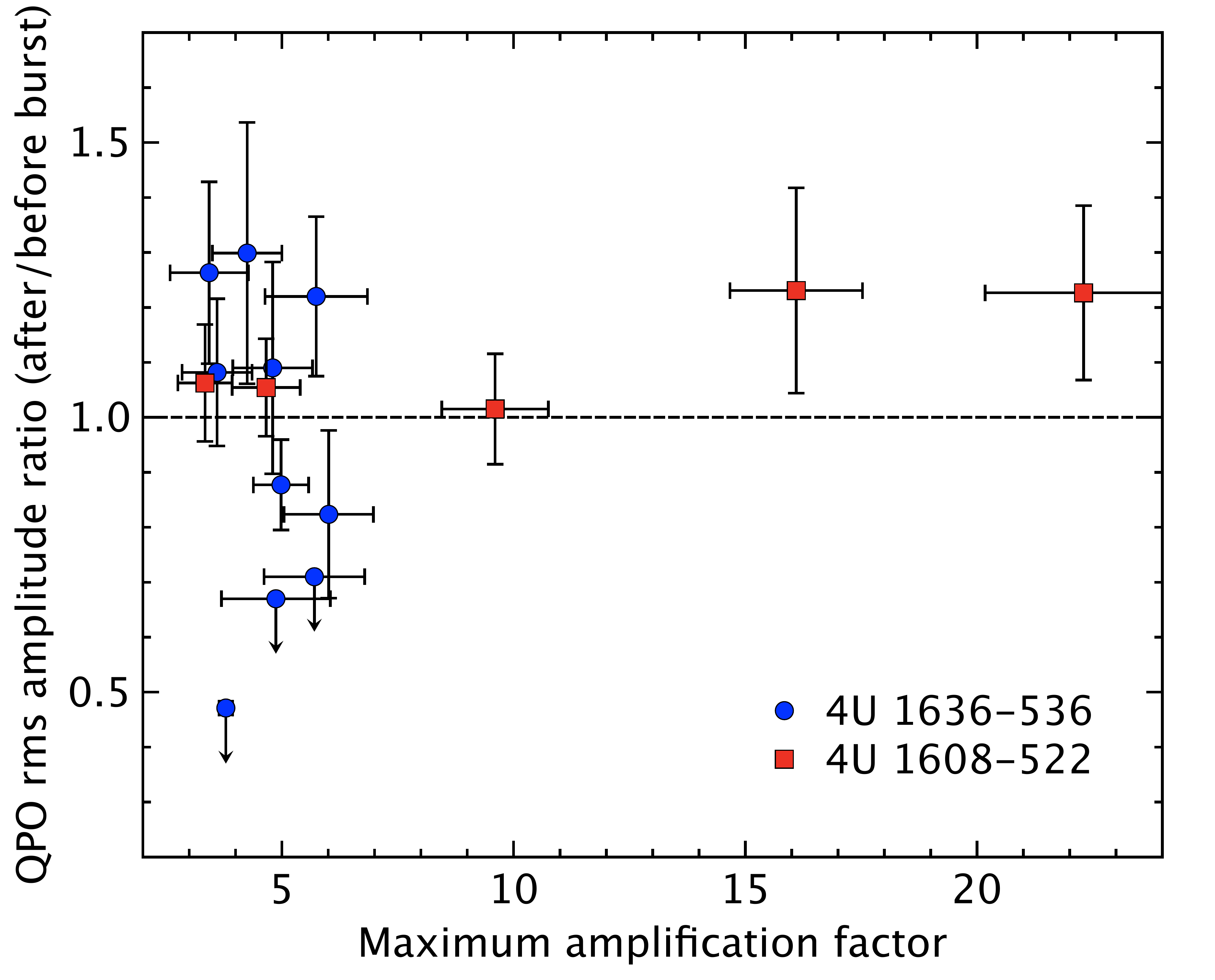} &
	\includegraphics[width = 0.4\linewidth]{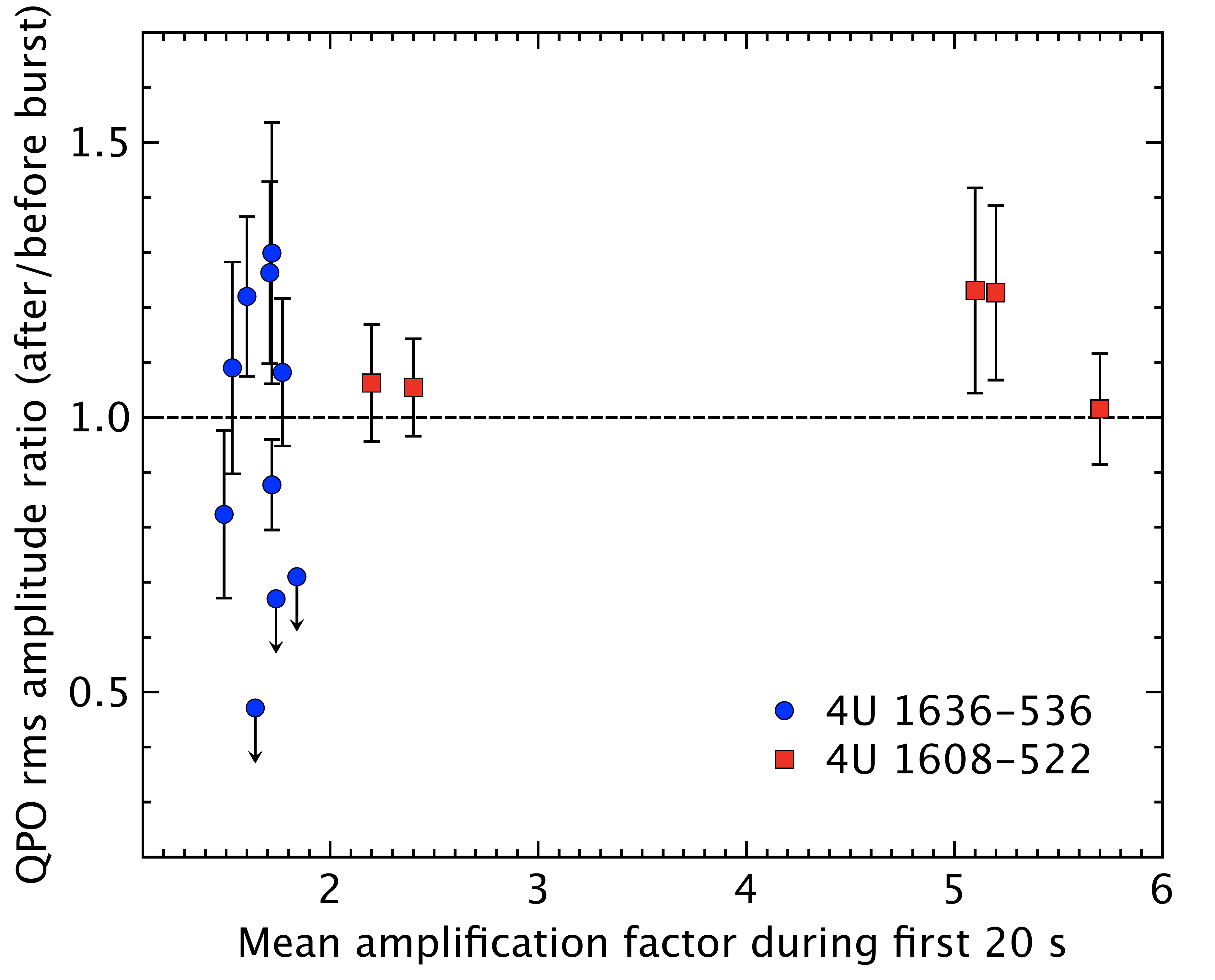}\\
	\end{tabular}
	\caption{Ratio of the kHz QPO rms amplitude after the burst over its value before the burst (see Table \ref{tab:para_fit}) as a function of the burst peak luminosity (upper left), burst total energy (upper right), maximum $f_a$ (lower left), and mean $f_a$ during the first 20~s (lower right) with 1$\sigma$ error bars.  The upper limits are 90\% level. The blue circles correspond to bursts of 4U 1636-536, the red squares to bursts of 4U 1608-522. When no significant signal was detected, upper limits were plotted (bursts 4, 22, and 23 of 4U 1636-536).}
	\label{fig:rms_fct_peak}
\end{figure*}
In the above analysis, we have shown that not all QPOs respond the same way to Type I X-ray bursts. Here we wish to investigate whether this can be related to changes in the burst parameters. In Fig. \ref{fig:rms_fct_peak} we plot the ratio of the rms amplitude of the QPO after and before the burst (see Table \ref{tab:para_fit} for more detail), against the peak luminosity and burst total energy. As can be seen, no clear trend is observed within the limited sample of the bursts considered in our analysis concerning the peak luminosity. Nonetheless, the burst showing a highly significant disappearance of the QPO (4) -- as well as the two bursts where the lack of QPO signal is marginally significant (22 and 23) -- has a peak luminosity that lies on the high part of the diagram. It is striking to see that burst 4 has an exceptionally large total energy which is due to its long duration. Unfortunately, extending our analysis to bursts observed after the \citet{Galloway:2008} catalog did not provide us with further useful observations and these trends could not be confirmed.

\begin{figure*}
	\centering
	\begin{tabular}{ccc}
	\resizebox{0.315\hsize}{!}{\includegraphics{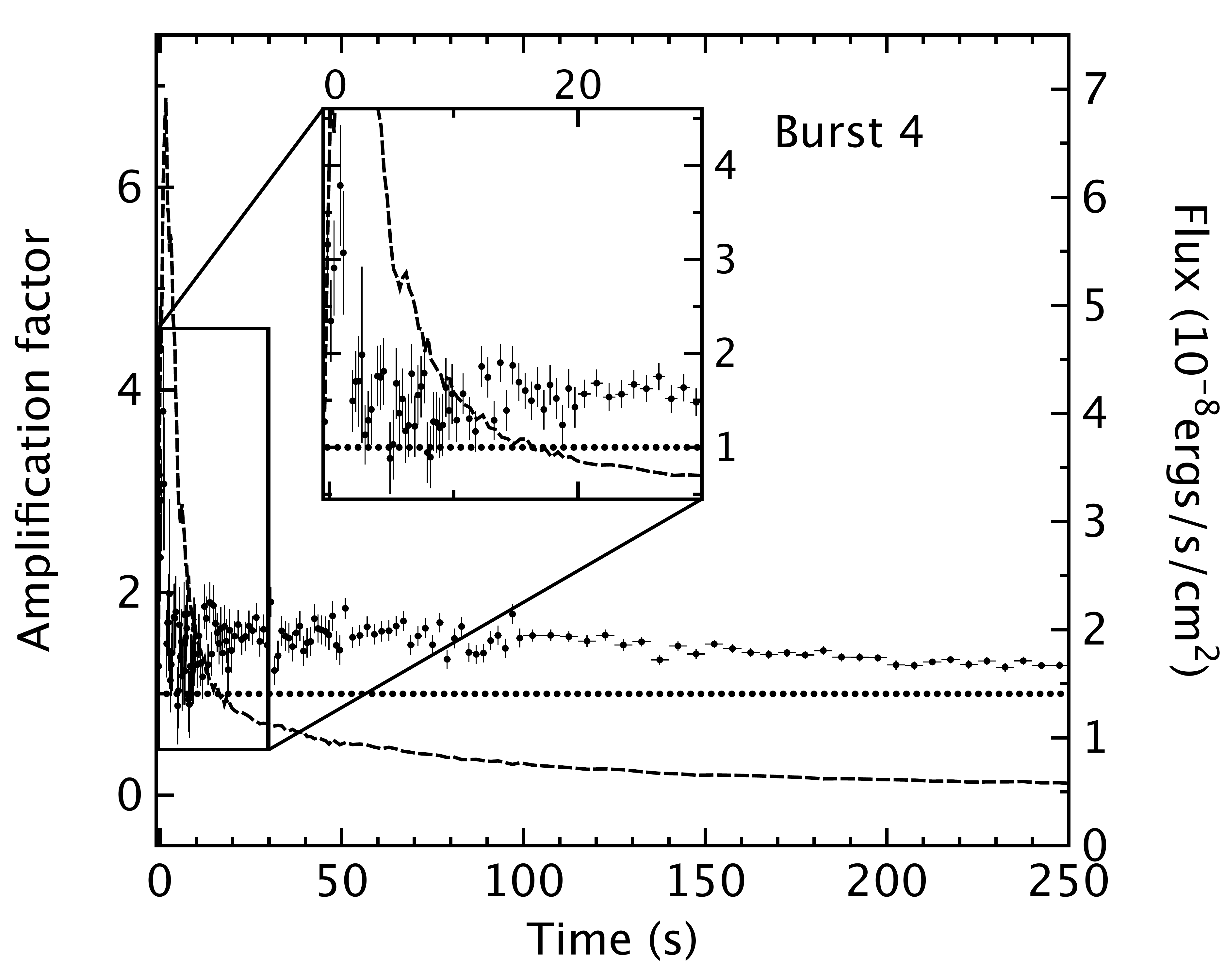}} &
	\resizebox{0.315\hsize}{!}{\includegraphics{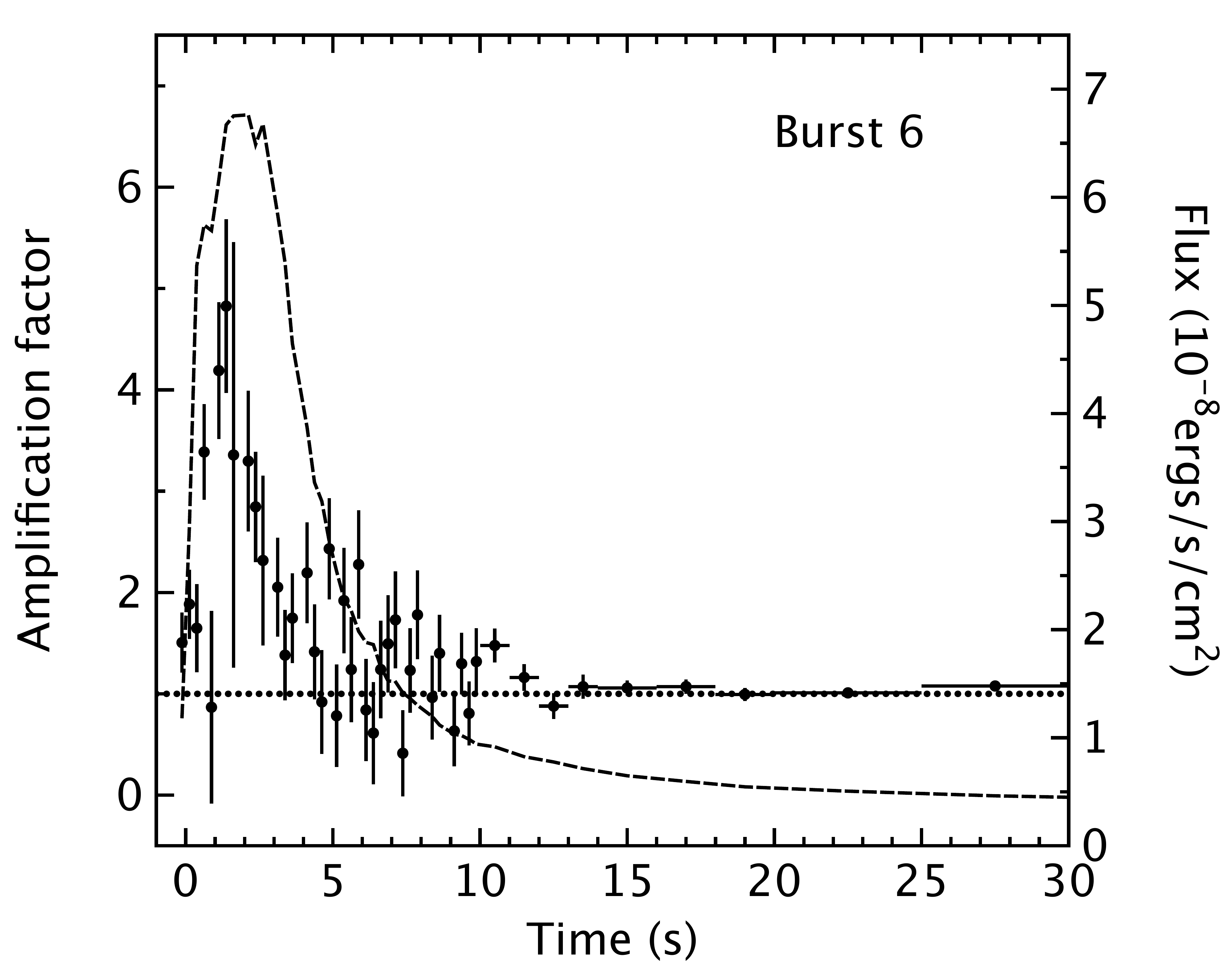}} &
	\resizebox{0.315\hsize}{!}{\includegraphics{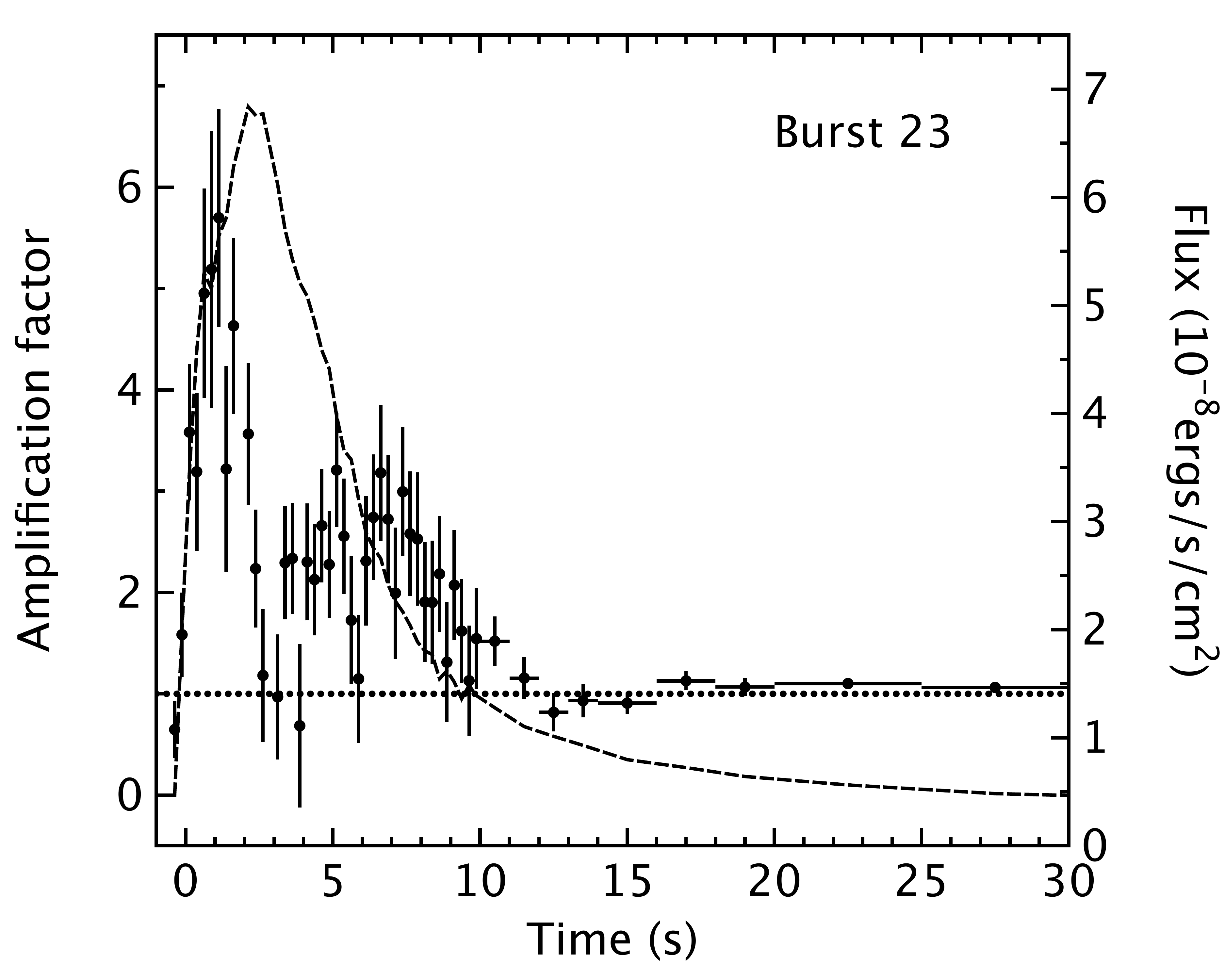}}\\
	\end{tabular}
	\caption{Amplification factor ($f_a$) profiles (points with 1$\sigma$ error bars) obtained from the spectral analysis of bursts 4 (left), 6 (middle), and 23 (right) of 4U 1636-536. For the longer burst 4, the first 30~s of the $f_a$ profile have been magnified for better comparison. The dashed lines correspond to the burst light curves, and the dotted horizontal lines to $f_a=1$.}
	\label{fig:fa_profiles}
\end{figure*}

We have also performed a spectral analysis of our burst sample following the recent work of \citet{Worpel:2013aa}, where the time resolved burst spectrum is fitted as the sum of an absorbed blackbody and a scalable persistent emission spectrum (with a scaling factor $f_a$), as fitted before the burst. Fixed hydrogen column densities of N$_H=0.36 \times 10^{21}$ \citep{Pandel:2008} and $1.1 \times 10^{21}$ cm$^{-2}$ \citep{Guver:2010aa} were used for 4U 1636-536 and 4U 1608-522, respectively. Spectra were fitted in the 2.5 -- 20 keV band using \texttt{xpsec} version 12.8. Burst spectra were extracted with variable time steps matching the burst dynamics. Two different models were used to fit the persistent emission spectrum: \texttt{wabs*(nthcomp+diskline)} for 4U 1636-536 and \texttt{wabs*(nthcomp+gaussian+bbodyrad)} for 4U 1608-522. As discussed by \citet{Worpel:2013aa}, what matters is to have a model that describes accurately the persistent emission before the bursts. Both models provide good and stable fits. Like \citet{Worpel:2013aa}, we found that $f_a$ increases during the peak phase (up to 20 or so, with typical values around 5), and remains significantly above 1 over the 20-s interval following the burst. An example of $f_a$ variations for three bursts is shown in Fig. \ref{fig:fa_profiles}. No correlation was found between the ratio of the QPO rms amplitude before and after the burst and $f_a$, neither its maximum value nor its mean value over the first 20-s interval (as shown in Fig. \ref{fig:rms_fct_peak}). Yet, it is interesting to note that with the exception of the long burst of 4U 1636-536 (Fig. \ref{fig:4_22_23}, left panel), $f_a$ has returned to 1 within 20~s. This value is commensurable with the recovery time inferred from the stacking of the 4U 1608-522 data. 

\section{Discussion}
\label{discussion}
High frequency QPOs are commonly believed to be generated close to the neutron star surface or in the innermost parts of the accretion disk. Similarly, Type I X-ray bursts are produced on the neutron star surface and radiate away enough energy to change the proprieties of the accretion flow. It is therefore natural to investigate the impact of a Type I X-ray burst on kHz QPOs. Our finding that there may be cases where the QPO is affected by the bursts on timescales of hundreds of seconds and cases where the QPO does not suffer from the bursts down to the shortest timescales that can be investigated by current data ($\sim$20~s) is clearly puzzling, especially because these two behaviors cannot be irrefutably connected in our limited sample to different properties of the bursts, such as the burst peak luminosity. 
We now discuss our results in the framework of the recent work by \citet{Worpel:2013aa}, claiming that the accretion rate is enhanced during Type I X-ray bursts.
	\subsection{A disk recession?}	
As accretion disks are optically thick in the radial direction, an enhanced accretion rate due to radiation torque can only be attained through disk depletion: $f_a > 1$ implies a receding disk. The Eddington-scaled accretion rate in our data sets is $\sim$0.12 and 0.04~$\dot{M}_{\text{Edd}}$ for 4U 1636-536 and 4U 1608-522, respectively, using a neutron star mass of 1.4~M$_{\odot}$. If we take conservative values for the mean $f_a$ measured during the 20~s following the burst onset of 1.5 and 2 for the two sources, this gives depleted masses of at least $\sim3\times10^{18}$ and $2\times10^{18}$~g. Using standard accretion disks as described in \citet{Shakura:1973} to model the mass distribution in the inner parts of the disk with an $\alpha$-parameter of 0.1, these masses correspond approximately to the mass contained between 7--50~$R_g$ for 4U 1636-536, and 7--30~$R_g$ for 4U 1608-522. It is important to note that the disk surface density scales as 1/$\alpha$ which is poorly constrained and that these masses are only rough estimates.

An estimate of the timescale at which the disk recovers its former geometry is given by the viscous time \citep[][Eq. 5.69]{Frank:2002},
\begin{equation}
t_{\text{visc}} \sim 3 \times 10^5 \alpha^{-4/5} \left(\dfrac{\dot{M}}{10^{16}\text{g/s}}\right)^{-3/10}\left(\dfrac{M}{\text{M}_{\odot}}\right)^{1/4}\left(\dfrac{R}{10^{10}\text{cm}}\right)^{5/4}~s,
\end{equation} 
with $R$ the distance to the neutron star center, $\dot{M}$ the accretion rate, and $M$ the neutron star mass. For $M = 1.4$ M$_{\odot}$, $\alpha = 0.1$, and with the appropriate accretion rates, we find $t_{\text{visc}} \sim 130$~s at 50~$r_g$ for 4U 1636-536 and $t_{\text{visc}} \sim 100$~s at 30~$r_g$ for 4U 1608-522. It is worth noting that increasing $\alpha$ gives shorter viscous times, but it also increases the radius matching the loss of mass and similar timescales are found. Within all the caveats of the above assumptions, it is interesting to note that these timescales match the nondetection gap of the two shorter bursts from 4U 1636-536, shown in Fig. \ref{fig:4_22_23}. We note however that for the longer burst of 4U 1636-536, $f_a$ is still above 1 while the QPO reappears (see Fig. \ref{fig:fa_profiles}), which argues against the idea that the disk is still truncated through depletion. This clearly suggests that caution should be used when using $f_a$ to derive the accretion rate during the burst. Some refinements in the \citet{Worpel:2013aa} model might be necessary such as considering modifications of the spectral shape of the persistent emission during the burst. If such high levels of disk depletion occur, one would expect at least the disk emission to be modified; for example, a change in the disk inner radius from 7~$r_g$ to 50~$r_g$ would decrease its inner temperature by a factor of 4.4,  assuming a dependency of $\propto R^{-3/4}$ as in \citet{Shakura:1973}. The timescales found are also a factor of 5-10 longer than the recovery time we inferred from 4U 1608-522, thus suggesting that an alternative to disk depletion should be considered.

\begin{figure}[!t]
	\resizebox{\hsize}{!}{\includegraphics{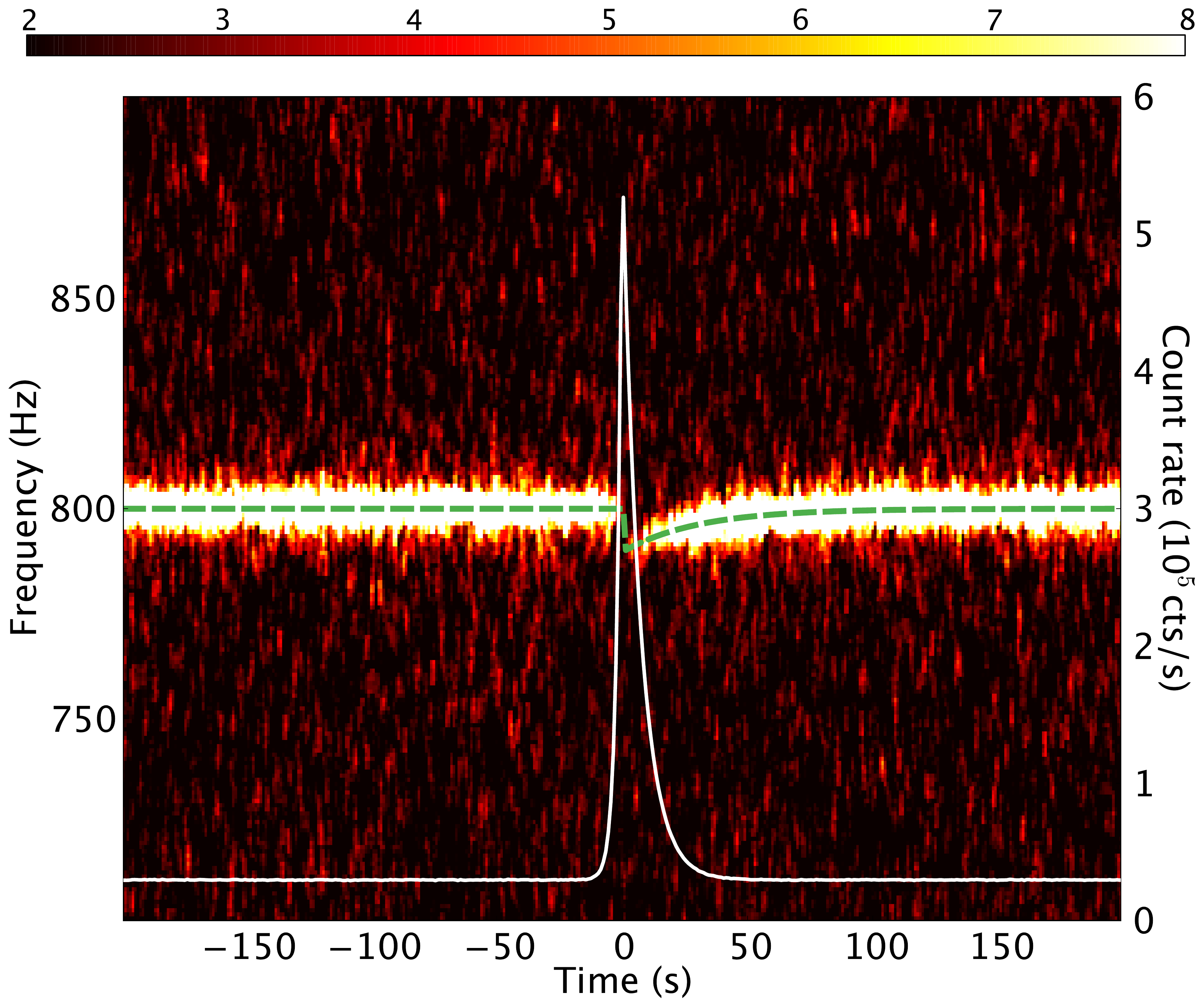}}
	\centering
	\caption{Simulated dynamical PDS containing a QPO at 800 Hz and a smooth frequency jump of 10 Hz initiated at the burst peak (the dashed green line gives the QPO frequency as a function of time). The QPO rms amplitude is 8\% and the QPO width is 3 Hz. The LOFT/LAD 3--30\,keV light curve of the Type I X-ray burst (peak at 18 times the persistent emission level, rise time of 2\,s, and decay time of 7\,s) is overplotted with a white line. The persistent count rate was estimated using 4U 1608-522 as the source. The image corresponds to a series of 1 s PDS plotted as a function of time, convolved with a 6 Hz and 2 s averaging kernel. Power is color coded. We note that the color scale is different from the one used for RXTE data: it saturates at a power level of 8 instead of 3.}
	\label{fig:LOFT}
\end{figure}
\subsection{Heating of the inner region?}

To reduce the viscous disk recovery time, only the innermost part of the accretion disk should be affected. At 10~$R_g$, the viscous time $\sim$18~s for 4U 1636-536 and $\sim$25~s for 4U 1608-522 with an $\alpha$-parameter of 0.1. One possibility could thus be disk irradiation by the burst photons, which is expected to produce significant heating, and might cause the disk to puff up \citep{Ballantyne:2005}. Leaving aside the two shorter bursts of 4U 1636-536 for which the QPO disappearance is marginally significant, such a mechanism could be consistent with the late reappearance of the QPO in the longer one (while the burst is still ongoing), but also with the short recovery times of the QPOs in 4U 1608-522. This might indicate that the modulated emission reappears only when the level of heating is sufficiently low. Unfortunately, current data cannot tell us what happens to the QPO within the first 20~s in 4U 1608-522, e.g., whether its amplitude drops or its coherence decreases. Hence, although that seems unlikely, we also cannot exclude that there is no effect of the burst on the QPO in this source.

Existing data are therefore insufficient to access timescales shorter than 20~s, while we would like to measure the QPO parameters as close as possible to the burst peak. Assuming that the QPO amplitude and width remain constant within the burst and only the QPO frequency varies, in Fig. \ref{fig:LOFT} we show that a new generation timing instrument like the LOFT/LAD \citep{Feroci:2012} would allow us to track the QPO frequency, even at the burst peak. Tracking the QPO frequency along the frequency drift assumed in the simulation requires an increase in effective area by at least one order of magnitude (15 in the case considered here\footnote{LOFT/LAD response files were downloaded from \url{http: //www.isdc.unige.ch/loft} for the simulation. We have assumed the goal effective area of 12~m$^2$ at 8~keV for the LAD. }). This simulation illustrates how such a large area timing instrument could deliver breakthrough observations on the burst-QPO interaction, providing at the same time insights on the location of the QPO modulated emission. 

	\subsection{Conclusions}
	
The interaction between the kHz QPO signals and Type-I X-ray bursts has been studied in two LMXBs: 4U 1636-536 and 4U 1608-522. Two types of behaviors of different timescales have been identified. We have found clear evidence of an interaction of the burst with the QPO emission during an exceptionally long burst (and a possible indication for two others) while for most of them, within the current data, the QPO does not seem to be affected by the burst. We have set an upper limit of 20~s for the recovery time of the QPO in 4U 1608-522. We have shown that a next generation timing mission providing an increase in effective area of at least an order of magnitude, such as LOFT, would be able to detect the QPO throughout the bursts and hence provide better constraints on the physics of the interaction of the burst emission and its surroundings. This would need to be complemented by theoretical work aimed at a better modeling of the burst-disk interaction.

\begin{acknowledgements} 

This research has made use of data obtained through the High Energy Astrophysics Science Archive Research Center On-line Service, provided by the NASA/Goddard Space Flight Center. We thank Duncan K. Galloway and Hauke Worpel for helpful conversations and suggestions. We also thank Mariano Mendez for a careful review of the paper, and comments which helped to improve its quality.

\end{acknowledgements}

\bibliographystyle{aa}
\bibliography{Biblio_these}

\begin{thebibliography}{30}
\expandafter\ifx\csname natexlab\endcsname\relax\def\natexlab#1{#1}\fi

\bibitem[{{Ballantyne} \& {Everett}(2005)}]{Ballantyne:2005}
{Ballantyne}, D.~R. \& {Everett}, J.~E. 2005, \apj, 626, 364

\bibitem[{{Ballantyne} \& {Strohmayer}(2004)}]{Ballantyne:2004}
{Ballantyne}, D.~R. \& {Strohmayer}, T.~E. 2004, \apjl, 602, L105

\bibitem[{{Barret}(2013)}]{Barret:2013}
{Barret}, D. 2013, \apj, 770, 9

\bibitem[{{Barret} {et~al.}(2005){Barret}, {Klu{\'z}niak}, {Olive}, {Paltani},
  \& {Skinner}}]{Barret:2005}
{Barret}, D., {Klu{\'z}niak}, W., {Olive}, J.~F., {Paltani}, S., \& {Skinner},
  G.~K. 2005, \mnras, 357, 1288

\bibitem[{{Barret} {et~al.}(2006){Barret}, {Olive}, \& {Miller}}]{Barret:2006}
{Barret}, D., {Olive}, J.-F., \& {Miller}, M.~C. 2006, \mnras, 370, 1140

\bibitem[{{Barret} \& {Vaughan}(2012)}]{Barret:2012aa}
{Barret}, D. \& {Vaughan}, S. 2012, \apj, 746, 131

\bibitem[{{Berger} {et~al.}(1996){Berger}, {van der Klis}, {van Paradijs},
  {Lewin}, {Lamb}, {Vaughan}, {Kuulkers}, {Augusteijn}, {Zhang}, {Marshall},
  {Swank}, {Lapidus}, {Lochner}, \& {Strohmayer}}]{Berger:1996}
{Berger}, M., {van der Klis}, M., {van Paradijs}, J., {et~al.} 1996, \apjl,
  469, L13

\bibitem[{{Chen} {et~al.}(2011){Chen}, {Zhang}, {Torres}, {Zhang}, {Li},
  {Kretschmar}, \& {Wang}}]{Chen:2011}
{Chen}, Y.-P., {Zhang}, S., {Torres}, D.~F., {et~al.} 2011, \aap, 534, A101

\bibitem[{{Cumming}(2004)}]{Cumming:2004aa}
{Cumming}, A. 2004, Nuclear Physics B Proceedings Supplements, 132, 435

\bibitem[{{Degenaar} {et~al.}(2013){Degenaar}, {Miller}, {Wijnands},
  {Altamirano}, \& {Fabian}}]{Degenaar:2013}
{Degenaar}, N., {Miller}, J.~M., {Wijnands}, R., {Altamirano}, D., \& {Fabian},
  A.~C. 2013, \apjl, 767, L37

\bibitem[{{Feroci} {et~al.}(2012){Feroci}, {Stella}, {van der Klis},
  {Courvoisier}, {Hernanz}, {Hudec}, {Santangelo}, {Walton}, {Zdziarski},
  {Barret}, {Belloni}, {Braga}, {Brandt}, {Budtz-J{\o}rgensen}, {Campana}, {den
  Herder}, {Huovelin}, {Israel}, {Pohl}, {Ray}, {Vacchi}, {Zane}, {Argan},
  {Attin{\`a}}, {Bertuccio}, {Bozzo}, {Campana}, {Chakrabarty}, {Costa}, {De
  Rosa}, {Del Monte}, {Di Cosimo}, {Donnarumma}, {Evangelista}, {Haas},
  {Jonker}, {Korpela}, {Labanti}, {Malcovati}, {Mignani}, {Muleri},
  {Rapisarda}, {Rashevsky}, {Rea}, {Rubini}, {Tenzer}, {Wilson-Hodge},
  {Winter}, {Wood}, {Zampa}, {Zampa}, {Abramowicz}, {Alpar}, {Altamirano},
  {Alvarez}, {Amati}, {Amoros}, {Antonelli}, {Artigue}, {Azzarello},
  {Bachetti}, {Baldazzi}, {Barbera}, {Barbieri}, {Basa}, {Baykal}, {Belmont},
  {Boirin}, {Bonvicini}, {Burderi}, {Bursa}, {Cabanac}, {Cackett}, {Caliandro},
  {Casella}, {Chaty}, {Chenevez}, {Coe}, {Collura}, {Corongiu}, {Covino},
  {Cusumano}, {D'Amico}, {Dall'Osso}, {De Martino}, {De Paris}, {Di Persio},
  {Di Salvo}, {Done}, {Dov{\v c}iak}, {Drago}, {Ertan}, {Fabiani}, {Falanga},
  {Fender}, {Ferrando}, {Della Monica Ferreira}, {Fraser}, {Frontera},
  {Fuschino}, {Galvez}, {Gandhi}, {Giommi}, {Godet}, {G{\"o}{\v g}{\"u}{\c s}},
  {Goldwurm}, {G{\"o}tz}, {Grassi}, {Guttridge}, {Hakala}, {Henri}, {Hermsen},
  {Horak}, {Hornstrup}, {in't Zand}, {Isern}, {Kalemci}, {Kanbach}, {Karas},
  {Kataria}, {Kennedy}, {Klochkov}, {Klu{\'z}niak}, {Kokkotas}, {Kreykenbohm},
  {Krolik}, {Kuiper}, {Kuvvetli}, {Kylafis}, {Lattimer}, {Lazzarotto}, {Leahy},
  {Lebrun}, {Lin}, {Lund}, {Maccarone}, {Malzac}, {Marisaldi}, {Martindale},
  {Mastropietro}, {McClintock}, {McHardy}, {Mendez}, {Mereghetti}, {Miller},
  {Mineo}, {Morelli}, {Morsink}, {Motch}, {Motta}, {Mu{\~n}oz-Darias},
  {Naletto}, {Neustroev}, {Nevalainen}, {Olive}, {Orio}, {Orlandini},
  {Orleanski}, {Ozel}, {Pacciani}, {Paltani}, {Papadakis}, {Papitto},
  {Patruno}, {Pellizzoni}, {Petr{\'a}{\v c}ek}, {Petri}, {Petrucci}, {Phlips},
  {Picolli}, {Possenti}, {Psaltis}, {Rambaud}, {Reig}, {Remillard},
  {Rodriguez}, {Romano}, {Romanova}, {Schanz}, {Schmid}, {Segreto}, {Shearer},
  {Smith}, {Smith}, {Soffitta}, {Stergioulas}, {Stolarski}, {Stuchlik},
  {Tiengo}, {Torres}, {T{\"o}r{\"o}k}, {Turolla}, {Uttley}, {Vaughan},
  {Vercellone}, {Waters}, {Watts}, {Wawrzaszek}, {Webb}, {Wilms}, {Zampieri},
  {Zezas}, \& {Ziolkowski}}]{Feroci:2012}
{Feroci}, M., {Stella}, L., {van der Klis}, M., {et~al.} 2012, Experimental
  Astronomy, 34, 415

\bibitem[{{Frank} {et~al.}(2002){Frank}, {King}, \& {Raine}}]{Frank:2002}
{Frank}, J., {King}, A., \& {Raine}, D.~J. 2002, {Accretion Power in
  Astrophysics: Third Edition}

\bibitem[{{Galloway} {et~al.}(2008){Galloway}, {Muno}, {Hartman}, {Psaltis}, \&
  {Chakrabarty}}]{Galloway:2008}
{Galloway}, D.~K., {Muno}, M.~P., {Hartman}, J.~M., {Psaltis}, D., \&
  {Chakrabarty}, D. 2008, \apjs, 179, 360

\bibitem[{{G{\"u}ver} {et~al.}(2010){G{\"u}ver}, {{\"O}zel}, {Cabrera-Lavers},
  \& {Wroblewski}}]{Guver:2010aa}
{G{\"u}ver}, T., {{\"O}zel}, F., {Cabrera-Lavers}, A., \& {Wroblewski}, P.
  2010, \apj, 712, 964

\bibitem[{{in't Zand} {et~al.}(2011){in't Zand}, {Galloway}, \&
  {Ballantyne}}]{int-Zand:2011}
{in't Zand}, J.~J.~M., {Galloway}, D.~K., \& {Ballantyne}, D.~R. 2011, \aap,
  525, A111

\bibitem[{{in't Zand} {et~al.}(2013){in't Zand}, {Galloway}, {Marshall},
  {Ballantyne}, {Jonker}, {Paerels}, {Palmer}, {Patruno}, \&
  {Weinberg}}]{int-Zand:2013aa}
{in't Zand}, J.~J.~M., {Galloway}, D.~K., {Marshall}, H.~L., {et~al.} 2013,
  \aap, 553, A83

\bibitem[{{Kuulkers} {et~al.}(2003){Kuulkers}, {den Hartog}, {in't Zand},
  {Verbunt}, {Harris}, \& {Cocchi}}]{Kuulkers:2003}
{Kuulkers}, E., {den Hartog}, P.~R., {in't Zand}, J.~J.~M., {et~al.} 2003,
  \aap, 399, 663

\bibitem[{{Leahy} {et~al.}(1983){Leahy}, {Darbro}, {Elsner}, {Weisskopf},
  {Kahn}, {Sutherland}, \& {Grindlay}}]{Leahy:1983}
{Leahy}, D.~A., {Darbro}, W., {Elsner}, R.~F., {et~al.} 1983, \apj, 266, 160

\bibitem[{{M{\'e}ndez} {et~al.}(1999){M{\'e}ndez}, {van der Klis}, {Ford},
  {Wijnands}, \& {van Paradijs}}]{Mendez:1999}
{M{\'e}ndez}, M., {van der Klis}, M., {Ford}, E.~C., {Wijnands}, R., \& {van
  Paradijs}, J. 1999, \apjl, 511, L49

\bibitem[{{Nakamura} {et~al.}(1989){Nakamura}, {Dotani}, {Inoue}, {Mitsuda},
  {Tanaka}, \& {Matsuoka}}]{Nakamura:1989}
{Nakamura}, N., {Dotani}, T., {Inoue}, H., {et~al.} 1989, \pasj, 41, 617

\bibitem[{{Pandel} {et~al.}(2008){Pandel}, {Kaaret}, \& {Corbel}}]{Pandel:2008}
{Pandel}, D., {Kaaret}, P., \& {Corbel}, S. 2008, \apj, 688, 1288

\bibitem[{{Serino} {et~al.}(2012){Serino}, {Mihara}, {Matsuoka}, {Nakahira},
  {Sugizaki}, {Ueda}, {Kawai}, \& {Ueno}}]{Serino:2012}
{Serino}, M., {Mihara}, T., {Matsuoka}, M., {et~al.} 2012, \pasj, 64, 91

\bibitem[{{Shakura} \& {Sunyaev}(1973)}]{Shakura:1973}
{Shakura}, N.~I. \& {Sunyaev}, R.~A. 1973, \aap, 24, 337

\bibitem[{{Strohmayer} \& {Bildsten}(2003)}]{Strohmayer:2003aa}
{Strohmayer}, T. \& {Bildsten}, L. 2003, ArXiv Astrophysics e-prints

\bibitem[{{Timmer} \& {Koenig}(1995)}]{Timmer:1995}
{Timmer}, J. \& {Koenig}, M. 1995, \aap, 300, 707

\bibitem[{{van der Klis}(1989)}]{van-der-Klis:1989}
{van der Klis}, M. 1989, in Timing Neutron Stars, ed. H.~{{\"O}gelman} \&
  E.~P.~J. {van den Heuvel}, 27

\bibitem[{{van der Klis}(2006)}]{van-der-Klis:2006}
{van der Klis}, M. 2006, {Rapid X-ray Variability}, 39--112

\bibitem[{{Walker}(1992)}]{Walker:1992}
{Walker}, M.~A. 1992, \apj, 385, 642

\bibitem[{{Worpel} {et~al.}(2013){Worpel}, {Galloway}, \&
  {Price}}]{Worpel:2013aa}
{Worpel}, H., {Galloway}, D.~K., \& {Price}, D.~J. 2013, \apj, 772, 94

\bibitem[{{Yu} {et~al.}(1999){Yu}, {Li}, {Zhang}, \& {Zhang}}]{Yu:1999}
{Yu}, W., {Li}, T.~P., {Zhang}, W., \& {Zhang}, S.~N. 1999, \apjl, 512, L35

\end{thebibliography}

\end{document}